\documentclass[journal, twocolumn]{IEEEtran}

\usepackage[utf8]{inputenc} 
\usepackage[T1]{fontenc}
\usepackage{url}
\usepackage{ifthen}
\usepackage{cite}
\usepackage[cmex10]{amsmath} 
\usepackage{amssymb}
\usepackage{graphicx}
\usepackage{bm}
\usepackage{bbm}
\usepackage{color}
\usepackage{mathtools}
\usepackage{balance}
\usepackage{caption} 

\newcommand{\lp}{\left(}
\newcommand{\rp}{\right)}
\newcommand{\lb}{\left[}
\newcommand{\rb}{\right]}
\newcommand{\lbp}{\left\{}
\newcommand{\rbp}{\right\}}
\newcommand{\lba}{\left\lvert}
\newcommand{\rba}{\right\rvert}

\newcommand{\mv}{\middle\vert}
\newcommand{\ul}{\underline}
\newcommand{\ol}{\overline}
\newcommand{\mcal}{\mathcal}

\newcommand{\msf}{\mathsf}

\newcommand{\ra}{\rightarrow}

\newcommand{\eqDef}{\triangleq}

\newcommand{\E}{\mathbb{E}}

\renewcommand{\Pr}{\mathbb{P}}
\newcommand{\Ber}{\mathrm{Ber}}

\newcommand{\Indc}[1]{\mathbbm{1}{\lbp #1\rbp}}

\newcommand{\KLD}[2]{{D}\left( #1\, \middle\Vert #2 \right)}
\newcommand{\kld}[2]{{d}\lp #1\, \middle\Vert #2 \rp}

\newcommand{\wSet}{\lb1:n\rb}
\newcommand{\mSet}{\lb0:M-1\rb}
\newcommand{\aSet}{\mathcal{A}}
\newcommand{\zSet}{\mathcal{Z}}
\newcommand{\info}{F}
\newcommand{\test}{\phi}

\newcommand{\bbeta}{ \boldsymbol{\beta}}

\newtheorem{theorem}{Theorem}
\newtheorem{assumption}{Assumption}
\newtheorem{remark}{Remark}
\newtheorem{example}{Example}
\newtheorem{fact}{Fact}
\newtheorem{lemma}{Lemma}
\newtheorem{definition}{Definition}
\newtheorem{corollary}{Corollary}
\newtheorem{proposition}{Proposition}

\allowdisplaybreaks

\begin{document}
\title{Tradeoffs among Action Taking Policies Matter in Active Sequential Multi-Hypothesis Testing: the Optimal Error Exponent Region}

\author{
Chia-Yu Hsu,~\IEEEmembership{Member,~IEEE} and I-Hsiang Wang,~\IEEEmembership{Member,~IEEE} 
\thanks{This work was supported by NSTC of Taiwan under Grant 111-2628-E-002-005-MY2 and 113-2628-E-002-022-MY4, and NTU under Grant 113L7764, 113L891404, and 113L900902. 
The material in this paper was presented in part at the 2024 IEEE International Symposium on Information Theory, July 2024.}%
\thanks{C.-Y Hsu was with the Graduate Institute of Communication Engineering, 
National Taiwan University, Taipei 10617, Taiwan. He is now with the Department of Electrical and Computer Engineering, University of Michigan, Ann Arbor (email: chiayuh@umich.edu).}
\thanks{I.-H. Wang is with the Department of Electrical Engineering and the Graduate Institute of Communication Engineering, National Taiwan University, Taipei 10617, Taiwan (email: ihwang@ntu.edu.tw).}
}

\maketitle

\begin{abstract}
Reliability of sequential hypothesis testing can be greatly improved when the decision maker is given the freedom to adaptively take an action that determines the distribution of the current collected sample. Such advantage of sampling adaptivity has been realized since Chernoff's seminal paper in 1959 \cite{Chernoff_59}. While a large body of works have explored and investigated the gain of adaptivity, in the general multiple-hypothesis setting, the fundamental limits of individual error probabilities have not been fully understood. In particular, in the asymptotic regime as the expected stopping time tends to infinity, the error exponents are only characterized in specific cases, such as that of the total error probability. In this paper, we consider a general setup of active sequential multiple-hypothesis testing where at each time slot, a temporally varying subset of data sources (out of a known set) emerges from which the decision maker can select to collect samples, subject to a family of expected selection budget constraints. The selection of sources, understood as the ``action'' at each time slot, is constrained in a predefined action space. At the end of each time slot, the decision maker either decides to make the inference on the $M$ hypotheses, or continues to observe the data sources for the next time slot. The optimal tradeoffs among $M(M-1)$ types of error exponents are characterized. A companion asymptotically optimal test that strikes the balance between exploration and exploitation is proposed to achieve any target error exponents within the region. To the best of our knowledge, this is the first time in the literature to identify such tradeoffs among error exponents in active sequential hypothesis testing, and it uncovers the tension among different action taking policies even in the basic setting of Chernoff \cite{Chernoff_59}.
\end{abstract}

\section{Introduction}\label{sec:intro}
Adaptivity in taking samples from different distributions is known to greatly enhance the performance of sequential statistical inference. The intuition is that different distributions have different levels of informativeness of the inference target, and the most informative one may vary with the ground truth. 
The decision maker builds up more confidence towards the ground truth as it collects more samples, so it should choose the most informative distribution at the moment accordingly. 
As for hypothesis testing, the paradigm is called active sequential hypothesis testing \cite{Chernoff_59, NaghshvarJavidi_13,NitinawaratAtia_13}, where the decision maker has the freedom to adaptively take an action that determines the distribution of the sample on the fly. 
It can be understood as the setting where the decision maker selects a particular data source out of multiple ones to collect a sample, each of which corresponds to a possible action. 
This paradigm finds extensive application in various fields such as spectrum sensing, crowdsourced data analytics, etc., and motivated from these applications, there have been extensions that incorporate a variety of expected selection budget constraints \cite{BaiGupta_17,LiLi_19,BaiKatewa_15,Vaidhiyan2015} and temporally changing available data sources \cite{LanWang_21}. 

Nevertheless, the fundamental limits of individual error probabilities have not been fully understood in existing works about active sequential multi-hypothesis testing \cite{Chernoff_59, NaghshvarJavidi_13, NitinawaratAtia_13, BaiGupta_17, BaiKatewa_15, LiLi_19, Vaidhiyan2015, LanWang_21}. The main focus of existing works was on minimizing costs related to the expected stopping time subject to constant constraints on certain kinds of error probabilities (typically weighted combinations of individual error probabilities). 
When the treatment pertains to the general non-asymptotic regime, it is usually difficult to characterize the optimal performance analytically, and hence the goal becomes developing efficient algorithms for finding the optimal selection policy numerically under a given framework. 
Meanwhile for the asymptotic regime, one can reformulate the problems in \cite{Chernoff_59, NaghshvarJavidi_13, NitinawaratAtia_13, Vaidhiyan2015} by investigating the optimal performance metric (total error probability, maximal error probability, or decision risk) under constraints on expected stopping time and resource budgets, as in \cite{LanWang_21} for binary hypothesis testing. It is not difficult to characterize the \emph{exponentially decaying rates (error exponents)} in the respective settings, and the analytical results provide insights on how action taking policies impact the fundamental performance. 
However, in the general $M$-ary multi-hypothesis setting, there are in total $M(M-1)$ types of error probabilities, and it remains unclear how the asymptotically optimal policies and inference strategies change when trading off one error probability with others.

Towards finer resolution of the fundamental performance limits in active sequential $M$-ary multi-hypothesis testing problems, in this work we focus on $M(M-1)$ types of \emph{individual} error probabilities in a general framework that covers those in the literature \cite{DragalinTartakovsky_99, Chernoff_59, NaghshvarJavidi_13, NitinawaratAtia_13, BaiGupta_17, BaiKatewa_15, LiLi_19, Vaidhiyan2015, LanWang_21} as special cases. In this framework, out of $n$ data sources in total, at each time slot a \emph{random} subset of them is revealed to the decision maker as the available sources. The distribution of this random subset is i.i.d. over time. The decision maker then takes an action (may be randomized) to select data sources from this available subset and collects samples from the selected ones. 
Note that the action taking policy (selection rule) is \emph{adaptive}, that is, it depends on the past observed samples. 
At the end of each time slot, according to the samples accumulated so far, the decision maker either determines to make an inference among the $M$ hypotheses or advances to the next time slot. 
Under a \emph{family} of expected selection budget constraints including the expected stopping time, the goal is to study the asymptote of individual error probabilities, in particular the exponentially decaying rates (error exponents), as all budget constraints scale to infinity with fixed ratios. 

Our primary contribution lies in the characterization of the optimal individual error exponent region, revealing the \emph{existence of tradeoffs among certain groups of individual error exponents}. Moreover, we prove a unified theorem that generalizes across diverse scenarios, including binary and multiple hypotheses testing, as well as settings with a single or multiple data sources with or without random availability. 
We show that the achievable error exponents can be represented as linear combinations of Kullback-Leibler (KL) divergences, each of which corresponds to a particular set of data sources' joint level of informativeness for distinguishing one hypothesis from another. The coefficients in the linear combinations correspond to the expected selection frequencies of data sources under the given policy and turn out to be the key design choices. In our characterization, these selection frequencies are indexed naturally by the taken action and the available subset of data sources. 
The optimal error exponent region is the union of these $M(M-1)$-tuples of error exponents over all possible such selection frequencies that meet the \emph{normalized} expected budget constraints (i.e., the ratios to the expected stopping time) and the distribution of the random available subset.
Leveraging the fact that these constraints are all affine, we further note that the optimal error exponent region is a convex set. 
The characterized region turns out to be a direct product of $M$ sub-regions each of which is in an $(M-1)$-dimensional space, 
and the \emph{tradeoff} only appears among the exponents that correspond to the same declared hypothesis against all other $M-1$ possible ground truth hypotheses.
To the best of our knowledge, such tradeoffs among $M(M-1)$ types of error exponents have not been identified before, and it uncovers the tension among different action taking policies even in the basic setting of Chernoff \cite{Chernoff_59}. 
The characterization of the optimal region also helps elucidate the gain of adaptivity in terms of error exponents, which has not been fully understood in the literature. To do so, we analyze the optimal achievable region of error exponents for \emph{non-adaptive} selection rules, that is, selection rules that are independent of the information collected up to the current time. It is then shown that the optimal achievable region of the adaptive setting subsumes that of the non-adaptive setting. Numerical results are also provided to visualize the gain. 

To characterize the optimal error exponent region, we divide the proof of our main theorem into two parts: the converse part and the achievability part. 
The converse part of our main theorem is proved by standard arguments that involves data processing inequality and Doob's Optional Stopping Theorem, similar to that in \cite{LanWang_21}. The more interesting part is achievability. 
The stopping and inference rules of our test mainly follow a family of multi-hypothesis sequential probability ratio tests (MSPRT) proposed in \cite{DragalinTartakovsky_99}, with thresholds designed judiciously to achieve the target error exponents and meet the expected stopping time constraint. For the action taking policy, since the constraints are all in expectation, it is natural to use a randomized one, and hence the design boils down to choosing the action taking \emph{probabilites}. 
To achieve a given tuple of error exponents in the region, the coefficients in the linear combinations of the KL divergences serve as the guidance for setting these action taking probabilities in the randomized policy. As suggested in the literature \cite{Chernoff_59, NaghshvarJavidi_13, NitinawaratAtia_13,LanWang_21}, one way is to \emph{exploit} the accumulated information and choose the action taking probabilities derived from the linear combinations that correspond to the hypothesis with the maximum likelihood. 
It turns out that similar to \cite{Chernoff_59}, such a ``pure-exploitation'' approach requires stronger assumptions on the data sources so that it can be shown to be asymptotically optimal. Inspired by \cite{Vaidhiyan2015}, in our proposed policy, on top of the ``pure-exploitation'' part, we further mix it with a carefully designed probability for ``exploration'' so that more relaxed assumptions can be set. 

\subsection*{Literature review}
The study of asymptotic fundamental limits of hypothesis testing problems could be traced back to Hoeffding \cite{Hoeffding_65} and Blahut \cite{Blahut_74}, where the optimal tradeoff between the type-I and the type-II error exponents was characterized for fixed-length binary hypothesis testing. 
For sequential binary hypothesis testing, based on Wald and Wolfowitz's seminal result \cite{WaldWolfowitz_48}, it can be shown (see e.g. \cite[Chapter 16.3]{polyanskiy2025information}) that under the expected stopping time constraints, the type-I and type-II error exponents can be \emph{maximized simultaneously}, thereby eliminating the tradeoff observed in the fixed-length setting and demonstrating the gain of sequentiality (variable-length sample size) in error exponents. 
For multiple hypothesis testing, the optimal individual error exponent region in the fixed-length setting was characterized by Tuncel \cite{Tuncel_05}, while for the the sequential setting, one can follow a similar recipe in \cite[Chapter 16.3]{polyanskiy2025information} and derive the optimal error exponent region based on Dragalin \textit{et al.}'s work \cite{DragalinTartakovsky_99}. It can then be shown that sequentiality also eliminates the tradeoffs among the individual error exponents. 

The focus of the works above is on a \emph{passive} setting for decision making, where the decision maker collects data from a single source throughout the entire process. An \emph{active} setting with multiple sources was introduced by Chernoff \cite{Chernoff_59}, where the decision maker can dynamically select data sources, either adaptively or non-adaptively, thereby influencing the distribution of the collected data samples. Following Chernoff's seminal work \cite{Chernoff_59}, various extensions have been made \cite{NaghshvarJavidi_13,NitinawaratAtia_13,BaiGupta_17,LiLi_19,BaiKatewa_15,Vaidhiyan2015}, and they all take a similar problem formulation as Chernoff \cite{Chernoff_59}: minimize a cost related to the expected stopping time subject to constant constraints on certain kinds of error probabilities, such as a constant total error probability \cite{NaghshvarJavidi_13}, a constant maximal error probability \cite{Chernoff_59, Vaidhiyan2015}, constant type-I and type-II error probabilities \cite{LiLi_19,BaiKatewa_15,BaiGupta_17}, and constant decision risks \cite{NitinawaratAtia_13}. 
Meanwhile, a comprehensive study by Naghshvar and Javidi \cite{NaghshvarJavidi_13_2} compared various settings, including fixed-length versus sequential (variable-length), passive versus active, and non-adaptive versus adaptive decision making. The problem formulation is also to minimize a cost related to the expected stopping time under a constant total error probability constraint \cite{NaghshvarJavidi_13_2}, and it is shown that combining sequentiality and adaptivity yields the most favorable performance. 

The findings of these existing works indicate that sequentiality eliminates tradeoffs among error exponents when reformulated as an error probability minimization problem subject to expected stopping time constraints, and that adaptivity further improves error exponents while still avoiding tradeoffs. 
Interestingly when it comes to individual error exponents, our result reveals that tradeoffs among certain groups of \emph{individual} error exponents \emph{may exist} even for sequential adaptive tests, a phenomenon that has not been observed in the literature of active sequential multi-hypothesis testing. 

\subsection*{Organization of this paper} 
The remainder of this paper is organized as follows. In Section~\ref{sec:formulation}, we give the problem formulation, including the statistical model of collected samples and available subset of data sources, the family of active sequential tests, and the expected budget constraints. 
In Section~\ref{sec:pre_results}, we present results in the most basic setting of our problem formulation, namely, the setting considered in Chernoff's seminal paper \cite{Chernoff_59}, to highlight the tradeoffs among individual error exponents. 
We also compare our results with the error exponent region of non-adaptive decision rules to see the gain of adaptivity, along with numerical examples. 
The gain of sequentiality is demonstrated via the comparison with Tuncel's result \cite{Tuncel_05}. 
Then in Section~\ref{sec:results}, we present our main theorem in the most general setup and give an overview of the proposed asymptotically optimal test. 
The details of the proposed asymptotically optimal test and the achievability proof of our main theorem are given in Section~\ref{sec:achievability}. 
The details of the converse proof are provided in Section~\ref{sec:converse}. 
Section~\ref{sec:proof_of_lemma1} provides the details of the proof of a key technical lemma in the achievability part. 
Finally, in Section~\ref{sec:conclusion}, we conclude this paper. 

\subsection*{Differences from the conference version \cite{HsuWang_ISIT24}}
The conference version of this work \cite{HsuWang_ISIT24} considered a restricted setup where there are just two simple expected budget constraints: \emph{individual} and \emph{total} budget constraints, and the collected samples at each time slot are assumed to be conditionally independent random variables given the current action. In contrast, in this work, we consider a family of expected selection budget constraints, namely \emph{linear} expected budget constraints in the remaining context, and relax the conditional-independence assumption. 

\subsection*{Notations} 
In this paper, the following notations are employed. For $a < b$ being two integers, $[a:b]$ denotes the set of integers $\{a,a+1,\ldots,b-1, b\}$. 
$\mathbb{R}_+$ denotes the set of non-negative real numbers. 
$\bigtimes_{m=0}^{M-1}A_{m}$ denotes the direct product over sets, i.e., $A_0\times A_2\times ...\times A_{M-1}$. $\text{Conv}(A)$ and $\text{extr}(A)$ denote as the convex hull of set $A$ and the collection of extreme points of set $A$, respectively. $\{x_j\}_{j=1}^{k}$ denotes the $k$-tuple $\{x_1, x_2,...,x_k\}$ where $x_j$ is any value indexed by $j\in\{1,2,...,k\}$ for any integer $k\ge2$.

\section{Problem Formulation}\label{sec:formulation}
There are $M$ hypotheses and $n$ data sources labeled by integers in $\mSet \eqDef \{0,1,2,\ldots,M-1\}$ and $\wSet \eqDef \{1,2,\ldots,n\}$, respectively, where $2\le M<\infty$, $1\le n<\infty$. 
If not specified, in this paper upper and lower case letters are used to denote random variables and the values of random variables, respectively. $\mathbb{R}_+$ denotes the set of non-negative reals.

\subsection{Statistical model}\label{subsec: stat_model}
The statistical model of the general framework considered in this work comprises two parts: the available subset of data sources and the samples drawn from the data sources. 

At each time instance $t\ge 1$, the temporally available subset of data source $Z_t$ (a random variable), following a probability mass function (PMF) $P_Z$ (known to the decision maker) in an i.i.d. manner over time, is revealed to the decision maker. 
Let $\zSet=\{z_1,z_2,\ldots,z_{\lvert\zSet\rvert}\}$ denote the support of $P_Z$, $\lvert\zSet\rvert<\infty$. Each $z\in\zSet$ consists of a subset of data sources, i.e., $z_i\subseteq\wSet\ \forall\,i\in\{1,2,\ldots,\lvert\zSet\rvert\}$. 
The decision maker can select data sources from this available subset to collect data. 

As for the collected samples across time slots, under hypothesis $\theta\in\mSet$, samples generated by $n$ data sources (if they are \emph{selected}) follow a \emph{joint} distribution $P_{X_1,X_2,\ldots,X_n;\theta}$ where $X_j$ denotes the random variable sampled from the source $j\in\wSet$. Moreover, samples drawn from a subset of data sources (if selected) follow the  corresponding distribution marginalized from $P_{X_1,X_2,\ldots,X_n;\theta}$. For example, under hypothesis $\theta\in\mSet$, if $\{1, 2\}\subseteq\wSet$ is the selected subset of data sources, then the samples drawn from these two sources follow the distribution $P_{X_1,X_2;\theta}$ marginalized from $P_{X_1,X_2,\ldots,X_n;\theta}$. Note that we assume the decision maker exactly knows the distribution $P_{X_1,X_2,\ldots,X_n;\theta}$ for any $\theta\in\mSet$. For notational convenience, $\Pr_\theta\lp\cdot\rp$ and $\E_\theta[\cdot]$ respectively denote the short-hand notations for ``the probability'' and ``the expectation'' given the ground truth hypothesis is $\theta\in\mSet$.

\subsection{Action taking, stopping time, and inference}\label{sec:action_and_test}
At each time instance $t\ge 1$, the decision maker selects a subset of data sources from the available subset of data sources $Z_t$ mentioned above. Such a selection is an \emph{action} in the context of active sequential hypothesis testing. To make it formal, let us first introduce the following notations followed by illustrative examples.
\begin{itemize}
\item Action space $\aSet$: 
We denote the collection of all possible subsets of sources that the decision maker can choose as the \emph{action space} $\aSet$, and it is revealed to the decision maker in the very beginning. In other words, $\aSet$ collects all the possible actions (which subset of sources to choose) that the decision maker can take. In this paper, the action space is assumed to be a finite set, that is, $\lvert\aSet\rvert<\infty$. We also assume that the empty set $\emptyset\in\aSet$, that is, the decision maker is allowed to skip and not taking any source at a time instance. Note that the action space is part of the problem formulation and is fixed throughout the entire process of sequential inference.

\item Action $A_t$ at time instance $t$: 
We denote the selected subset of data sources as $A_t$, the action taken by the decision maker at time $t\ge 1$. In other words, $A_t\subseteq\wSet$ and $A_t\in\aSet$. Note that the actual subset of sources from which the decision maker eventually collect data samples is the \emph{intersection} of the selected subset $A_t$ and the temporarily available subset $Z_t$, that is , $A_t\cap Z_t$. 
\end{itemize}

\begin{example}
Cast in our formulation, the action space that appears in the literature \cite{LiLi_19,BaiKatewa_15, NaghshvarJavidi_13,Chernoff_59,NitinawaratAtia_13,Vaidhiyan2015,LanWang_21} is $\aSet=\{\emptyset, \{1\}, \ldots, \{n\}\}$, that is, the decision maker can select at most one data source at each time slot. In the existing works \cite{LiLi_19,BaiKatewa_15, NaghshvarJavidi_13,Chernoff_59,NitinawaratAtia_13,Vaidhiyan2015}, the entire set of $n$ sources are always available, and hence $Z_t = \wSet$ all the time. In \cite{LanWang_21}, the cardinality of $Z_t$ is always $1$. 
In general, each action in the action space can be an arbitrary \emph{subset} of $\wSet$, for example, $\aSet = \{\emptyset, \{1,2\}, \{1,3,5\}\}$, and the available subset of source $Z_t$ may vary from time to time, for example, $Z_1 = \{1,4,5\}$ and $Z_2 = \{3,4\}$. So suppose $A_2 = \{1,3,5\}$, then the subset of sources from which the decision maker can collect samples at time $t=2$ is $A_2\cap Z_2 = \{3\}$. 
\end{example}

The action taking process can be randomized, that is, $A_t$ at each time slot $t\geq1$ is a random subset, following a distribution determined by the decision maker. This distribution is the action taking policy and is \emph{adaptive} in general, that is, at each time instance $t\geq1$, the probability of taking $A_t\in\aSet$ is a function of the data (comprising samples, actions, and available sets) observed until time slot $t-1$. 

After taking an action at each time slot $t$, the decision maker collects a set of samples $\{X_{t,j}\,\vert\,j\in A_t\cap Z_t\}$ following the corresponding marginal distribution, where each $X_{t,j}$ is the sample generated by the data source $j$ that is selected, i.e., in the intersection of the action $A_t$ and the temporary available set $Z_t$. In short, $\forall\,t\geq1$, given $A_t=a\in\aSet$ and $Z_t=z\in\zSet$, the collected samples are represented as the random tuple $\mathbf{X}_t\eqDef\{X_{t,j}\,\vert\,j\in a\cap z\}$ following the distribution marginalized from $P_{X_1,X_2,\ldots,X_n;\theta}$, $P_{\{ X_{j} | j\in a\cap z\};\theta}$, under hypothesis $\theta\in\mSet$, $\forall\,a\in\aSet$, $\forall\,z\in\zSet$. For notational convenience, we denote 
\[
P^{a, z}_{\theta}\eqDef P_{\{ X_{j} | j\in a\cap z\};\theta}. 
\]
We assume given the actions, the random tuple $\mathbf{X}_t$ is conditionally independent (given the current action) across time slots. 
After receiving the tuple $\mathbf{X}_t$ at time slot $t\ge1$, the decision maker either stops and makes an inference or proceeds to the next time slot. The time that the decision maker stops is the \emph{stopping time}, denoted as $\tau\in\mathbb{N}$, and the inferred result is denoted as $\phi\in\mSet$. 

The procedures at the decision maker is summarized in the definition below. 
\begin{definition}[Active sequential test]\label{def:decision_maker}
An active sequential test in the proposed framework comprises the following two parts: at each time instance $t\geq1$, 
\begin{itemize}
\item The distribution that the randomized action $A_t$ follows is a function of $\info_{t-1}$ and $Z_t$, where
\[
\info_t \eqDef \{\{\mathbf{X}_s\}_{s=1}^{t}, \{A_s\}_{s=1}^{t}, \{Z_s\}_{s=1}^{t}\}\quad \forall\,t\geq1.
\]
By convention, $\info_0 = \emptyset$.
\item $\tau$ and $\phi$ are both functions of $\info_t$, $\forall\,t\ge1$. 
\end{itemize}
\end{definition}

As for the performance of the overall decision making, this work focuses on the individual error probabilities: $\forall\,\theta\in\mSet\setminus \{m\}, \forall\,m\in\mSet$,  
\[
    \pi_{m|\theta} \eqDef \Pr_{\theta}\lp \test = m\rp,
\]
the probability of mistaking $\theta$ for $m$. Note that there are in total $M(M-1)$ individual error probabilities.

\subsection{Budget constraints and error exponents}\label{subsec:exponents}
To formulate the budget constraints, let us first introduce the expected total number of times that source $j\in\wSet$ gets selected under hypothesis $\theta\in\mSet$:
\[
B^\theta_j \eqDef \E_\theta\lb\sum_{t=1}^{\tau}\Indc{j\in A_t\cap Z_t}\rb.
\]
To capture the various expected costs in the process of sequential decision making, let us then introduce a countable set of \emph{linear} functions $\big\{b_i \,\big\vert\, i\in \mathcal{I} \big\}$, $\mcal{I}$ is a countable index set, 
where each linear function
\[
b_i: \mathbb{R}^n_+\rightarrow\mathbb{R}_+, \ (B^\theta_1, B^\theta_2, \ldots, B^\theta_n) \mapsto b_i(B^\theta_1, B^\theta_2, \ldots, B^\theta_n)
\]
takes the expected total number of times that each source gets selected as the input and outputs a linear cost. In other words, we assume that the cost is a linear function of the total number of times that each source gets selected. Here $\mathbb{R}_+$ denotes the set of non-negative real numbers. Note that $b_i(\mathbf{0})=0$ for each $i\in\mcal{I}$. 

Let $\big\{U_i \,\big\vert\, i\in \mathcal{I} \big\}$ denote a set of non-negative real numbers, and each $U_i\in\mathbb{R}_+$ stands for the budget constraint on the cost $b_i$. 
To this end, we say that a decision maker ``satisfies the budget constraints'' if
\begin{equation}\label{eq:linear_exp_budg_constr}
b_i(B^\theta_1, B^\theta_2, \ldots, B^\theta_n)\leq U_i\quad\forall\,i\in\mathcal{I},\ \theta\in\mSet.
\end{equation}
Note that $\big\{U_i \,\big\vert\, i\in \mathcal{I} \big\}$ is a set of parameters in the problem formulation and given to the decision maker in the very beginning. For convenience, in this paper, we call Eq.~\eqref{eq:linear_exp_budg_constr} the ``\emph{linear} \emph{expected} budget constraints.''

\begin{remark}
Our formulation of budget constraints generalizes those in the literature. For example, in \cite{LiLi_19, BaiKatewa_15}, \emph{individual} expected budget constraints were considered, where $\mathcal{I}=\lbp 1,\ldots,n\rbp$, $b_i(\cdot)=\langle \cdot, \bm{e}_i\rangle$, $\forall\,i\in\mathcal{I}$, and each $\bm{e}_i\in\mathbb{R}^n_+$ is an unit vector with $1$ at the $i$-th entry and the rest being all $0$'s. Moreover, \cite{LanWang_21} considered \emph{total} expected budget constraints, where $\mathcal{I}=\{1\}$, $b_1(\cdot)=\langle \cdot, \bm{1}\rangle$, and $\bm{1}\in\mathbb{R}^n_+$ denotes the all-$1$ vector. Our formulation allows more general linear combinations of $(B^\theta_1, B^\theta_2, \ldots, B^\theta_n)$ than those in the previous works, taking into the account that accessing different sources might incur different costs.
\end{remark}

The expected stopping time constraints are defined as usual:
\[
\E_\theta[\tau]\leq T\quad\forall\,\theta\in\mSet, \label{eq:expected_stopping_time_cons}
\]
where $T\geq1$ stands for the constraint on the expected stopping time and it is also a parameter in the problem formulation and given to the decision maker in the very beginning

Our focus is on the fundamental limits of the performance in an asymptotic regime as $U_i\,, \forall\,i\in\mathcal{I}$ and $T$ all tend to infinity with fixed ratios
\[
r_i \eqDef \frac{U_i}{T} \geq 0\quad \forall\, i\in\mathcal{I}.
\]
The \emph{achievable} individual \emph{error exponent} 
\[
e_{m|\theta}, \forall\,\theta\in\mSet\setminus\{m\}, \forall\,m\in\mSet
\] 
is defined as follows.

\begin{definition}[Error exponents]\label{def:errorexponent}
A given $M(M-1)$-tuple 
\begin{equation}\label{eq:tuple}
\Big\{ e_{m\vert\theta} \,\Big\vert\, m\in\mSet, \theta\in\mSet\setminus\{m\}\Big\}
\end{equation}
is \emph{achievable} if there exists an active sequential test (comprising action taking, stopping time, and inference) satisfying the constraints \eqref{eq:linear_exp_budg_constr} and \eqref{eq:expected_stopping_time_cons} such that each of the $M(M-1)$ individual error probabilities \emph{vanishes} and satisfies 
\[
\begin{aligned}
&\liminf_{T\to\infty} \frac{1}{T} \log\frac{1}{\pi_{m|\theta}} \ge e_{m|\theta}\\
& \forall\,\theta\in\mSet\setminus\{m\}, \forall\,m\in\mSet.
\end{aligned}
\]
When the context is clear, we drop the indices and use $\{e_{m\vert\theta}\}$ to denote the $M(M-1)$-tuple in \eqref{eq:tuple}. 
\end{definition}

Our goal is to characterize the optimal error exponent \emph{region}, the collection of all achievable tuples.

\section{Main results in Chernoff's Setup}\label{sec:pre_results}
Before we give the full characterization of the optimal individual error exponent region in the most general setup, in this section we present a simplified version when the setup is specialized to that studied in Chernoff's paper \cite{Chernoff_59} and related works \cite{NaghshvarJavidi_13,NitinawaratAtia_13}. 
With the simplified version of our main result, we illustrate the tradeoffs among individual error exponents, make comparison with existing works, discuss the gain of adaptivity from the point of view of error exponents, and provide numerical examples. After the key insights and novelties are delivered and explained, the main technical result will follow in the next section. 

\subsection{Tradeoffs among individual error exponents}
To proceed, let us first recall the basic setup considered in \cite{Chernoff_59,NaghshvarJavidi_13,NitinawaratAtia_13}, where all data sources are available at each time instance, but only one data source can be selected at each time slot, and no budget constraints are imposed. In our notations, it corresponds to the following:
\[
\begin{aligned}
\zSet &=\lbp\wSet\rbp,& Z_t &= \wSet\ \text{with probability $1$},\\
\mcal{I} &= \emptyset,&
\aSet &=\lbp \emptyset, \{1\},\{2\},...\{n\}\rbp.
\end{aligned}
\]
Notably, since there is no budget constraint ($\mcal{I} = \emptyset$) and the only constraint is on the expected stopping time, the optimal action taking policy will choose $A_t = \emptyset$ with probability $0$. 

Under this setup, since there is no need to describe random availability and the selection is constrained to a single source, the notations in the statistical model can be simplified. Let us specialize the formulation to this setup and introduce corresponding simplified notations to better present the result. Note that they are used only in this section. In particular, for any hypothesis $m\in\mSet$ and data source $j\in\wSet$, if the true hypothesis is $m$, the data source $j$ (if it is selected) generates samples according to the distribution 
\[
P_{m, j}\equiv P_{X_j;m}, 
\]
the marginal distribution of data source $j$ under hypothesis $m$.

Before presenting the optimal exponent region, let us first provide an intuitive argument on a simple example to illustrate why the tradeoff among individual error exponents exists and how it looks like. 
Suppose there are just two data sources ($n=2$). First note that the error exponents depend largely on the \emph{selection frequency} (normalized by the expected stopping time constraint $T$) of the sources get selected. Let $\beta_1$ and $\beta_2$ denote the selection frequencies of source $1$ and $2$ respectively ($\beta_1,\beta_2\ge 0, \beta_1+\beta_2 = 1$). The achievable error exponents should be those satisfying
\[
\begin{aligned}
&e_{m|\theta}\leq\beta_1\KLD{P_{m,1}}{P_{\theta,1}} + \beta_2\KLD{P_{m,2}}{P_{\theta,2}}\\
&\forall\,m,\theta\in\mSet,\ m\neq\theta,
\end{aligned}
\]
and the right-hand-side could be interpreted as the average level of distinguishability of hypothesis $m$ from $\theta$ over the two data sources. 
The decision maker may choose the best selection frequencies to maximize the individual error exponent. However, the maximizer may be different across different $m,\theta$. 
For example, suppose data source $1$ is the most informative for distinguishing hypothesis $0$ from hypothesis $1$, that is, the KL divergence of $P_{0, 1}$ from $P_{1, 1}$ is larger than that of $P_{0, 2}$ from $P_{1, 2}$, that is, $\KLD{P_{0,1}}{P_{1,1}} > \KLD{P_{0,2}}{P_{1,2}}$. 
Meanwhile, data source $2$ is the most informative for distinguishing hypothesis $1$ from hypothesis $0$, that is, $\KLD{P_{1,1}}{P_{0,1}} < \KLD{P_{1,2}}{P_{0,2}}$. Then, to maximize $e_{0|1}$, one would choose data source $1$ over source $2$, that is, $(\beta_1,\beta_2) = (1,0)$, and to maximize $e_{1|0}$, one would do the other way around, that is, $(\beta_1,\beta_2) = (0,1)$. 
To this end, it seems impossible to achieve the best individual error exponents \emph{simultaneously}. Interestingly, contrary to the intuition, in the binary hypothesis setting, both $e_{0|1} = \KLD{P_{0,1}}{P_{1,1}}$ and $e_{1|0} = \KLD{P_{1,2}}{P_{0,2}}$ can be achieved simultaneously by the following \emph{adaptive} approach proposed by Chernoff \cite{Chernoff_59}: at each time instance, the decision maker first compute the likelihood functions of the two hypotheses based on the samples collected so far, and it chooses source $1$ if hypothesis $0$ is more likely, and source $2$ is hypothesis $2$ is more likely. 

Chernoff's approach for the general $M$-ary hypothesis setting can be described as follows \cite{Chernoff_59}: the decision maker can design $M$ selection frequency pairs $\big\{ (\beta^m_{1}, \beta^m_{2}) \big\}_{m=0}^{M-1}$, understood as $M$ different selection policies. If the likelihood of hypothesis $m\in\mSet$ is the largest, that is, the maximum likelihood estimate $\hat{\theta}_{\text{MLE}} = m$, it selects the data sources according to the corresponding frequency pair $\bm{\beta}^m \equiv (\beta^m_{1}, \beta^m_{2})$, indexed by the hypothesis $m$. However, unlike in the binary hypothesis testing setting where there is a single optimal $\bm{\beta}^m$ for each $m \in \{0,1\}$ (in the above example, the optimal $\bm{\beta}^0 = (1,0)$ and the optimal $\bm{\beta}^1 = (0,1)$), there may not exist such an optimal $\bm{\beta}^m$. The reason is that there are multiple individual error exponents correspond to the same $m$. For example, consider hypothesis $m=0$. Again, suppose data source $1$ is the most informative for distinguishing hypothesis $0$ from hypothesis $1$, that is, $\KLD{P_{0,1}}{P_{1,1}} > \KLD{P_{0,2}}{P_{1,2}}$. Meanwhile, data source $2$ is the most informative for distinguishing hypothesis $0$ from hypothesis $2$, that is, $\KLD{P_{0,1}}{P_{2,1}} < \KLD{P_{0,2}}{P_{2,2}}$. Then, to maximize $e_{0|1}$, one would choose data source $1$ over source $2$, that is, $\bm{\beta}^0 = (1,0)$, and to maximize $e_{0|2}$, one would do the other way around, that is, $\bm{\beta}^0 = (0,1)$. This time, the tension in choosing $\bm{\beta}^0$ immediately shows up, and it is hence impossible to achieve the best individual error exponents $e_{0|1}$ amd $e_{0|2}$ \emph{simultaneously}. Note that this tension cannot be mitigated by Chernoff's approach, as the two individual error exponents corresponds to the same $m$.

The above discussion illustrates the structure of tradeoffs among the individual error exponents: there is no tradeoff among $M$ error exponent $(M-1)$-tuples
\begin{equation}\label{eq:err_expt_tuples}
\begin{aligned}
&\big(e_{0|1},e_{0|2},\ldots,e_{0|M-1}\big),\   
\big(e_{1|0},e_{1|2},\ldots,e_{1|M-1}\big),\\ 
&\ldots,\ \big(e_{M-1|0},e_{M-1|2},\ldots,e_{M-1|M-2}\big),
\end{aligned}
\end{equation}
while tradeoff exists among $M-1$ individual error exponents $\big\{e_{m|\theta} \,\big\vert\, \theta \in \mSet\setminus\{m\} \big\}$ for each $m\in\mSet$. 
The following theorem summarizes the optimal error exponent region for Chernoff's setup with general number of sources.

\begin{theorem}[The optimal error exponent region in Chernoff's setup]
\label{thm:chernoff}
For any $n\geq1$ and $M\geq2$, the optimal achievable error exponent region for individual error probabilities in Chernoff's setup \cite{Chernoff_59} is given by
\begin{equation}\label{eq:err_expnt_region_chernoff}
\zeta=\bigtimes\limits_{m=0}^{M-1}\,\lbp\bigcup\limits_{\bbeta^m\in\mathcal{P}_n}\zeta_m(\bbeta^m)\rbp, 
\end{equation}
where $\mathcal{P}_n$ is the $n$-dimensional probability simplex, 
\begin{equation}\label{eq:selection_freq_tuple_chernoff}
\bbeta^m\eqDef\big(\beta^m_1,\beta^m_2,\ldots,\beta^m_n\big) \in \mathcal{P}_n
\end{equation} 
denotes a tuple of selection frequencies of $n$ data sources, and $\forall\,m\in\mSet$, the region 
$\zeta_m(\bbeta^m)$ is defined on the top of the next page.
\begin{figure*}[!t]
\normalsize
\begin{align}
\notag	&\zeta_m(\bbeta^m)\eqDef \lbp \big\{ e_{m\vert\theta} \,\big\vert\, \theta\in\mSet\setminus\{m\}\big\}\,\mv\, e_{m\vert\theta}\leq e^*_{m\vert\theta}(\bbeta^m)\quad \forall\,\theta\in\mSet\setminus\{m\}\rbp, \\
	&e^*_{m\vert\theta}(\bbeta^m)\eqDef\sum_{j=1}^{n}\beta^m_{j}\KLD{P_{m,j}}{P_{\theta,j}}\quad\forall\,\theta\in\mSet\setminus\{m\}. \label{eq:exponents_bd}
\end{align}
\hrulefill
\vspace*{4pt}
\end{figure*}
\end{theorem}

Theorem~\ref{thm:chernoff} is a special case of the general version (Theorem~\ref{thm:softErrorExponent} in the next section). Note that the direct product in \eqref{eq:err_expnt_region_chernoff} arises because the tradeoff only appears among the $M-1$ individual error exponents in $\bigcup_{\bbeta^m\in\mathcal{P}_n}\zeta_m(\bbeta^m)$ for each $m\in\mSet$. In words, for each declared hypothesis $m$, one may not find an optimal $\bbeta^m\in\mathcal{P}_n$ to maximize all $M-1$ bounds $\big\{ e^*_{m|\theta}(\bbeta^m)\,\big\vert\, \theta\in\mSet\setminus\{m\}\big\}$ simultaneously, leading to the tradeoff among these $M-1$ individual error exponents.

Theorem~\ref{thm:chernoff} also recovers the asymptotic limit of many different performance metrics in the literature. 
For example, \cite{Chernoff_59,NitinawaratAtia_13,NaghshvarJavidi_13} focused on a coarser performance metric, namely, the decision risk of declaring the decision as hypothesis $m$:
\begin{equation}\label{eq:decision_risk_def}
\mathrm{R}_m \eqDef \sum_{\theta\in\mSet\setminus\{m\}}\rho_\theta \pi_{m|\theta}\quad \forall\,m\in\mSet,
\end{equation}
where $\rho_\theta$ is the prior probability for hypothesis $\theta\in\mSet$.
The optimal exponentially decaying rates of the decision risks, denoted as $\{\gamma_{m} \mid m\in\mSet\}$, can then be characterized from our main result as a simple corollary given below.

\begin{corollary}[Exponents of decision risks]\label{cor:Decision_Risk_Exponent}
The optimal exponent region of the decision risks consists of non-negative $M$-tuples $(\gamma_0,...,\gamma_{M-1})$ satisfying
\[
\gamma_m \le \max_{\bbeta^m\in\mathcal{P}^n}\min_{\theta\in\mSet\setminus\{m\}}e^*_{m|\theta}(\bbeta^m)\quad\forall\,m\in\mSet,
\]
where $\{e^*_{m|\theta}(\bbeta^m)\}$ are defined in \eqref{eq:exponents_bd}.
\end{corollary}
Note that Corollary~\ref{cor:Decision_Risk_Exponent} can also be derived from the respective results in \cite{Chernoff_59,NitinawaratAtia_13,NaghshvarJavidi_13}. Clearly there are no tradeoffs among these $\gamma_m$'s due to the product form shown in \eqref{eq:err_expnt_region_chernoff}.

Meanwhile, suppose we take the error probabilities of individual hypotheses 
\[
\pi_\theta \eqDef \sum_{m\in\mSet\setminus\{\theta\}}\pi_{m|\theta}\quad \forall\,\theta\in\mSet 
\]
as performance metrics. We then denote the achievable exponentially decaying rates of the error probability $\pi_\theta$ as $e_\theta$ for any $\theta\in\mSet$. The following corollary of Theorem~\ref{thm:chernoff} is also immediate.
\begin{corollary}[Exponents of error probabilities of individual hypotheses]\label{cor:Error_Probability_Exponent}
The optimal exponent region of the error probabilities of individual hypotheses consists of non-negative $M$-tuples $(e_0,...,e_{M-1})$ satisfying
\[
e_\theta = \min_{m\in\mSet\setminus\{\theta\}} \{e_{m|\theta}\}
\quad\forall\,\theta\in\mSet
\]
for some $M(M-1)$-tuples $\big\{e_{m | \theta}\, \big\vert\, m\in\mSet\setminus\{\theta\}\big\} \in \zeta$, where $\zeta$ is given in \eqref{eq:err_expnt_region_chernoff}.
\end{corollary}
In contrast to the decision risks, there exist tradeoffs among the $M$ error exponents for error probabilities of individual hypotheses, and this has not been studied in the literature \cite{Chernoff_59,NitinawaratAtia_13,NaghshvarJavidi_13}.

\subsection{Gain of sequentiality and adaptivity}
In the following, we illustrate the gain of sequentiality and adaptivity by comparing the error exponent regions in the non-sequential non-adaptive \cite{Tuncel_05}, sequential non-adaptive \cite{DragalinTartakovsky_99}, and sequential adaptive settings. 
For fair comparison, we extend the results in the first two settings to the case where non-adaptive sampling from multiple data sources is allowed. 
First, we demonstrate that the optimal error exponent region of the sequential adaptive setting contains that of the sequential non-adaptive setting, which in turn contains that of the non-sequential non-adaptive setting. Then, we provide a numerical example to illustrate the gain of sequentiality and adaptivity for the case $M=3, n=2$. For the sake of illustration, as in Corollary~\ref{cor:Error_Probability_Exponent}, we take the error probabilities of individual hypotheses as performance metrics. 

In the non-sequential non-adaptive setting\cite{Tuncel_05}, we allow the decision maker to collect $T$ data samples independently from multiple data sources. The sampling proportions are specified by a fixed $\bbeta \in \mathcal{P}_n$, the $n$-dimensional probability simplex, chosen prior to data collection. For instance, if there are two data sources and $\bbeta = [0.3, 0.7]$, the decision maker will draw $0.3T$ samples from the first source and $0.7T$ samples from the second source. 
By the independence among samples and the non-adaptivity of sampling, it is straightforward to derive the optimal error exponent region based on Tuncel's result \cite{Tuncel_05}, as shown below: let $\zeta^{\text{Tuncel}}$ denote the optimal region, then
\[
\zeta^{\text{Tuncel}} = \bigcup_{\bbeta:\bbeta\in\mathcal{P}_n}\zeta^{\text{Tuncel}}(\bbeta)
\]
where $\zeta^{\text{Tuncel}}(\bbeta)$ is defined on the top of the next page.
\begin{figure*}[!t]
\normalsize
\begin{equation}
\zeta^{\text{Tuncel}}(\bbeta)\eqDef\lbp 
\big\{e_{m|\theta} \,\big\vert\, m,\theta\in\mSet,m\ne\theta\big\}\, \mv\, 
\begin{array}{l}\forall\, P_{j}\in \mcal{P}^{(j)},\ j\in[1:n],\ \exists\,\theta\in[0:M-1]\ \text{such that}\\
e_{m|\theta}\leq\sum_{j=1}^{n}\beta_{j}\KLD{P_{j}}{P_{\theta,j}}\quad \forall\, m\in[0:M-1]\setminus\{\theta\}
\end{array}
\rbp, \label{eq:err_expnt_region_tuncel}
\end{equation}
\hfill where $\mcal{P}^{(j)}$ denotes the set of all possible distributions of the alphabet of the $j$-th source, $j = 1,2,\ldots,n$. 
\par
\hrulefill
\end{figure*}

In the sequential non-adaptive setting, the decision maker selects an action $A_t$ at each time slot $t\ge1$ independently of the collected information $F_{t-1}$, where $A_t$ and $F_{t-1}$ are defined in Definition~\ref{def:decision_maker}. Due to the lack of adaptivity, the decision maker can no longer leverage Chernoff's approach to eliminate the tradeoffs among the $M$ error exponent $(M-1)$-tuples in \eqref{eq:err_expt_tuples}. Hence, the direct product in the achievable individual error exponent region \eqref{eq:err_expnt_region_chernoff} disappears, and it becomes
\begin{equation}\label{eq:err_expnt_region_nonadaptive}
    \zeta^{\text{NA}}=\bigcup_{\bbeta:\, \bbeta\in\mathcal{P}_n}\zeta^\text{NA}(\bbeta),
\end{equation}
where $\mathcal{P}_n$ is the $n$-dimensional probability simplex and $\zeta^\text{NA}(\bbeta)$ is defined on the top of the next page.
\begin{figure*}[!t]
\normalsize
\begin{align}
\notag    &\zeta^\text{NA}(\bbeta)\eqDef\big\{\{e_{m\vert\theta}\}\,\big\vert\, e_{m\vert\theta}\leq e^*_{m\vert\theta}(\bbeta)\quad \forall\,\theta\in\mSet,\ \forall \,m\in\mSet\setminus\{\theta\} \big\},\\
    &e^*_{m\vert\theta}(\bbeta)\eqDef\sum_{j=1}^{n}\beta_{j}\KLD{P_{m,j}}{P_{\theta,j}}\quad \forall\,\theta\in\mSet,\ \forall \,m\in\mSet\setminus\{\theta\} \quad \text{(the same as 
    \eqref{eq:exponents_bd})}.\label{eq:max_e_nonadp}
\end{align}
\hrulefill
\vspace*{4pt}
\end{figure*}
Here the superscript $(\cdot)^{\text{NA}}$ in the notation means ``non-adaptive.''

Comparing $\zeta^{\text{Tuncel}}$ and $\zeta^{\text{NA}}$, it can be immediately noted that $\zeta^{\text{Tuncel}} \subseteq \zeta^{\text{NA}}$, 
since for any given $\bbeta\in\mathcal{P}_n$, the largest possible $e_{m|\theta}$ in \eqref{eq:max_e_nonadp} can be shown to be in $e^*_{m\vert\theta}(\bbeta)$ in \eqref{eq:max_e_nonadp} using a similar argument as in \cite{Tuncel_05}. 
Meanwhile, comparing $\zeta^{\text{NA}}$ and the region $\zeta$ characterized in Theorem~\ref{thm:chernoff} (the sequential adaptive setting), 
it can be observed that 
\[
\zeta^{\text{NA}}(\bbeta) = \bigtimes\limits_{m=0}^{M-1}\zeta_m(\bbeta),
\]
and hence $\zeta^{\text{NA}} \subseteq \zeta$. 
As a result, we have $\zeta^{\text{Tuncel}} \subseteq \zeta^{\text{NA}}\subseteq\zeta$, demonstrating the gain of sequentiality and adaptivity.

To illustrate the gain of sequentiality and adaptivity, we adopt the following numerical example with two data sources ($n=2$) and three hypotheses ($M=3$). Let us consider the following $6$ ternary PMF's that correspond to the two data sources and three hypothesis respectively:
\[
\begin{array}{ll}
    P_{0,1} = [0.9\,,0.07\,,0.03], & P_{0,2} = [0.78\,,0.17\,,0.05];\\
    P_{1,1} = [0.12\,, 0.83\,, 0.05],& P_{1,2} = [0.04\,, 0.79\,, 0.17];\\
    P_{2,1} = [0.05\,, 0.1\,, 0.85],& P_{2,2} = [0.15\,, 0.05\,, 0.8].
\end{array}    
\]
Moreover, for visualization, we consider total error probabilities of individual hypotheses, that is,
\[
\pi_\theta =\sum_{m\in\mSet\setminus\{\theta\}}\pi_{m|\theta}\quad \forall\,\theta\in\mSet,
\]
for the three settings. Specifically, we focus on visualizing the achievable error exponent corresponding to each $\pi_\theta$ for $\theta\in\mSet$, where each exponent is denoted as $e_\theta$ for $\theta\in\mSet$. The resulting region for the adaptive setting is characterized in Corollary~\ref{cor:Error_Probability_Exponent}. 
Similarly, the corresponding regions for the non-sequential non-adaptive and sequential non-adaptive settings can be derived from~\eqref{eq:err_expnt_region_tuncel} and~\eqref{eq:err_expnt_region_nonadaptive}, respectively. 
We omit the description of these regions for brevity. Instead, below we only plot them for the considered numerical example. 
Since a three-dimensional plot of the error-exponent region is difficult to interpret, we instead illustrate three two-dimensional slices: the $(e_0, e_1)$ tradeoff under a fixed $e_2$, the $(e_0, e_2)$ tradeoff under a fixed $e_1$, and the $(e_1, e_2)$ tradeoff under a fixed $e_0$. 

\begin{figure*}[!t]
\includegraphics[width=\linewidth]{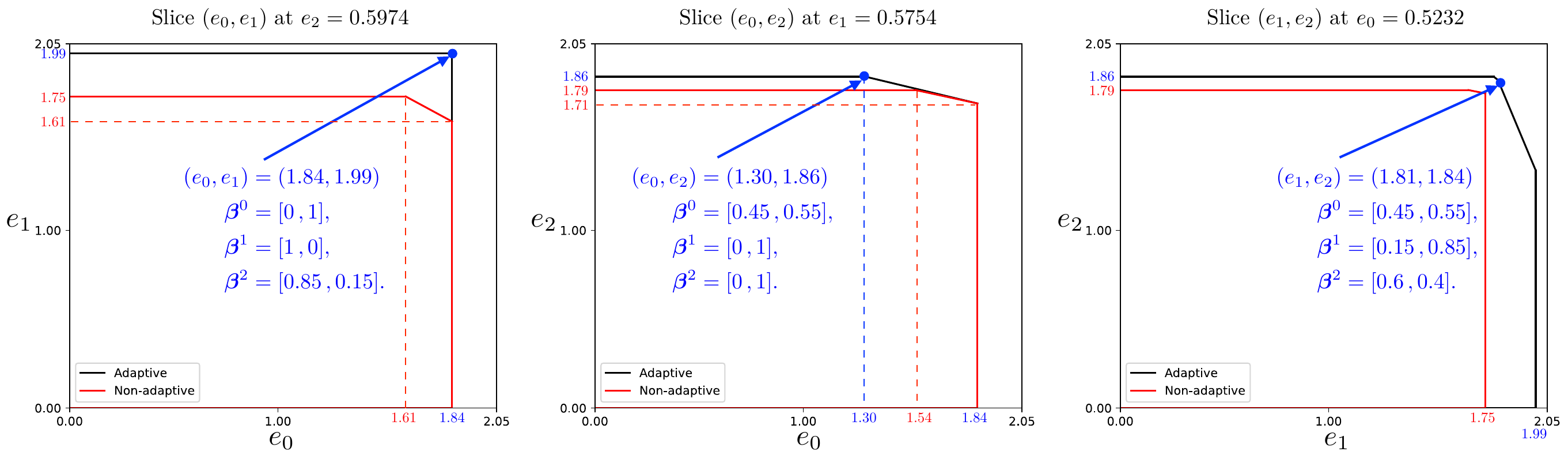}
\centering
\caption{Gain of adaptivity.}
\label{fig:gain_of_adaptivity}
\hrulefill
\vspace*{4pt}
\end{figure*}

Firstly, we show the gain of adaptivity in Figure~\ref{fig:gain_of_adaptivity}. The slice of the adaptive region ($\zeta$) and the slice of the non-adaptive region ($\zeta^{\text{NA}}$) are depicted by the black curve and the red curve, respectively. One can easily notice the non-adaptive region is strictly contained in the adaptive one. Moreover, in each slice, we highlight a blue dot representing an achievable error exponent tuple in the adaptive region that cannot be attained by any non-adaptive scheme, thereby demonstrating the advantage of being adaptive.

\begin{figure*}[!t]
\includegraphics[width=\linewidth]{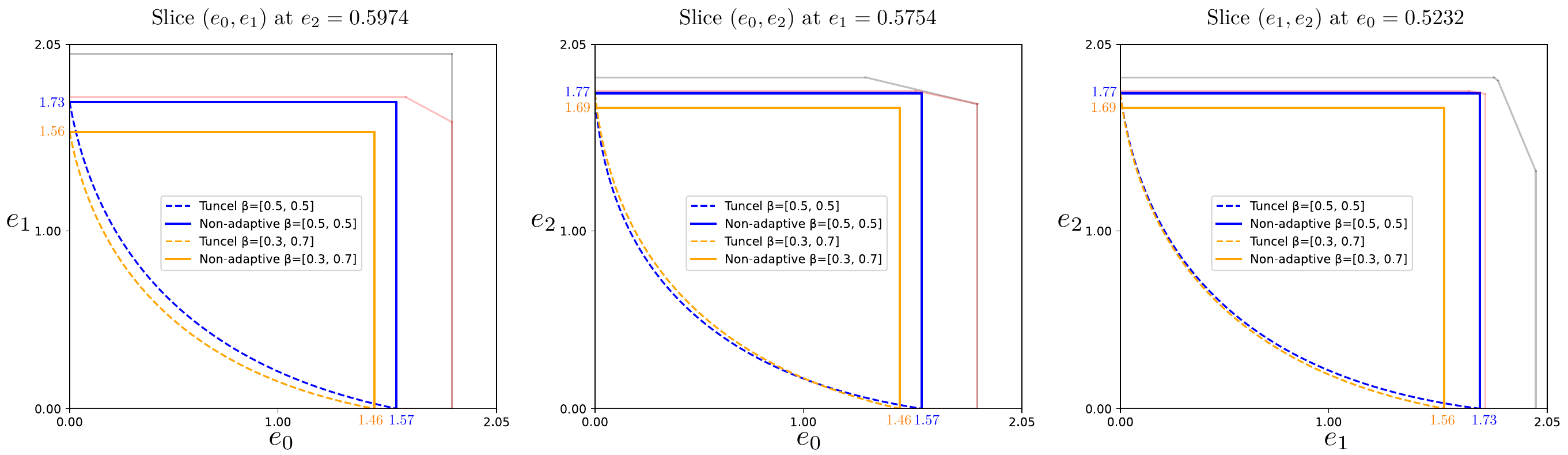}
\centering
\caption{Gain of sequentiality. The shaded lines denote the adaptive and non-adaptive regions in Fig.~\ref{fig:gain_of_adaptivity}.}
\label{fig:gain_of_sequentiality}
\hrulefill
\vspace*{4pt}
\end{figure*}

Secondly, we illustrate in Figure~\ref{fig:gain_of_sequentiality} the gain of sequentiality (in the absence of adaptivity). Since the region $\zeta^{\text{Tuncel}}$ is non-convex and difficult to plot, we draw slices of $\zeta^{\text{NA}}(\bbeta)$ and $\zeta^{\text{Tuncel}}(\bbeta)$ under two sets of $\bbeta$. In both cases, the solid curve corresponding to a slice of $\zeta^{\text{NA}}(\bbeta)$ encloses a larger region than the slice of $\zeta^{\text{Tuncel}}(\bbeta)$ (no tradeoff in the slice of $\zeta^{\text{NA}}(\bbeta)$), thereby demonstrating the advantage of sequential decision making over non-sequential strategies.

Let us conclude the section with two remarks that further highlight the benefits of adaptivity in different scenarios.
\begin{remark}[Gain of adaptivity when $M=2$]
In the binary hypothesis setting ($M=2$), \cite{LanWang_21} demonstrates that under the sequential adaptive setting, the optimal achievable $e_{0|1}$ and $e_{1|0}$ can be maximized simultaneously without any tradeoff. However, from the structure of $\zeta^{\text{NA}}$, it is evident that a tradeoff still exists between the optimal achievable $e_{0|1}$ and $e_{1|0}$ in the sequential non-adaptive setting due to the introduction of $\bbeta$.
\end{remark}

\begin{remark}[Gain of adaptivity for decision risks]
For the decision risks defined in \eqref{eq:decision_risk_def}, Corollary~\ref{cor:Decision_Risk_Exponent} implies that there is no tradeoff among the exponents of decision risks in the sequential adaptive setting. In contrast, in the sequential non-adaptive setting, such tradeoff do exits, which can be deduced from the structure of $\zeta^{\text{NA}}$ and the fact that the achievable exponents of the decision risks take the following form: $\gamma_m = \min_{\theta\in\mSet\setminus\{m\}}e_{m|\theta}$ for each $m\in\mSet$, where $\{e_{m|\theta}\}\in\zeta^{\text{NA}}$.
\end{remark}

\section{Main results in the General Setup}\label{sec:results}
In this section, we first present the optimal error exponent region for the general setup described in Section~\ref{sec:formulation}. 
Compared to the vanilla setup in Chernoff's work \cite{Chernoff_59}, the generalized setup further incorporates temporal variation of available data sources, the possibility to access multiple sources at each time slot, and general linear budget constraints. Nevertheless, the structure of the main theorem (Theorem~\ref{thm:softErrorExponent}) is similar to that of Theorem~\ref{thm:chernoff}. Accordingly, we interpret Theorem~\ref{thm:softErrorExponent} using the same insights as those obtained in Section~\ref{sec:pre_results} for Theorem~\ref{thm:chernoff}. We then discuss about the novelty and implications of the main theorem, emphasizing its contributions compared with existing literatures. Finally, we close this section with an overview of our proposed asymptotically optimal scheme. 

\subsection{Tradeoffs among individual error exponents}\label{sec:main_thm}
For the characterization of the optimal error exponents, the following two technical assumptions are imposed. 
\begin{assumption}
\label{a:oneStepLLR}
There exists $L<\infty$ such that
\[
\sup_{m\in\mSet}\sup_{\theta\in\mSet\setminus\{m\}}\sup_{\mathbf{x}\in\mcal{X}_\theta} \lba \log\frac{P_{X_1,\ldots,X_n;\theta}(\mathbf{x})}{P_{X_1,\ldots,X_n;m}(\mathbf{x})}\rba \le L,
\]
where $\mcal{X}_\theta$ denotes the support of the joint distribution $P_{X_1,\ldots,X_n;\theta}$, $\theta\in\mSet$. 
\end{assumption}

\begin{assumption}
\label{a:discrimination_1}
    For any two hypotheses $m,\theta\in\mSet$, $\theta\neq m$, 
\[
\exists\, z\in\zSet,\ \exists\, a\in\aSet\quad \text{such that}\quad \KLD{P^{a,z}_{m}}{P^{a,z}_{\theta}}>0.
\]
\end{assumption}

Assumption~\ref{a:oneStepLLR} is used in our achievability and converse proof and commonly made in the literature \cite{NaghshvarJavidi_13,Chernoff_59,NitinawaratAtia_13,Vaidhiyan2015,LanWang_21}. Assumption~\ref{a:discrimination_1} is needed for proving the achievability, and further discussion will be made in Section~\ref{sec:achievability}. Roughly speaking, it implies that for any two different hypotheses, there exists a feasible set of data sources that can distinguish one from the other and vice versa. Note that these two assumptions are also needed for Theorem~\ref{thm:chernoff}. 
However, to better serve the purpose of Section~\ref{sec:pre_results}, we do not state them explicitly there.

Before formally stating the theorem, let us first note that when generalizing the result from Chernoff's setup to the setup in Section~\ref{sec:formulation}, the definition of the selection frequency tuple and its constraint given in \eqref{eq:selection_freq_tuple_chernoff} inevitably require modification and extension. First of all, since the decision maker is faced with a temporally varying set of available data sources $Z_t$ (governed by $P_Z$) and it can select a subset of sources $A_t$ (recall $A_t\in\aSet$), one should focus on the selection frequency of each ``action'' $a\in\aSet$ when the available set of sources is $z\in\zSet$. Collecting all of them, with slight abuse of notation, we shall now use  
\[
\bbeta \eqDef\Big\{ \beta_{a,z} \,\Big\vert\, a\in\aSet, z\in\zSet\Big\} \in \mathbb{R}^{|\aSet||\zSet|}_+
\]
to denote a selection frequency tuple, where each component $\beta_{a,z}$ stands for the frequency that the selected set of data sources is $a$ and the available set of sources is $z$. 

Since we consider budget constraints and temporarily available data sources, unlike Chernoff's setup where the selection frequency tuple $\bbeta$ is only required to lie in the probability simplex $\mathcal{P}_n$, additional constraints on $\bbeta$ that correspond to budget and temporal availability emerge naturally. 
For each data source $j\in\wSet$, define
\begin{equation}\label{eq:select_freq_indv_src}
\omega_j(\bbeta) \eqDef \sum_{a\in\aSet}\sum_{z\in\zSet}\beta_{a,z}\Indc{j\in a\cap z},
\end{equation}
standing for the frequency with which source $j$ is selected during the decision process. Hence, multiplying \eqref{eq:select_freq_indv_src} by $T$ gives the \emph{expected} number of times that data source $j$ is selected in the decision process. Consequently, the budget constraints \eqref{eq:linear_exp_budg_constr} becomes 
\begin{equation}\label{eq:constraints_budget}
b_i\lp \omega_1(\bbeta),\ldots,\omega_n(\bbeta) \rp\le r_i\quad\forall\,i\in\mathcal{I}.
\end{equation}
Also note that $\sum_{a\in\aSet}\beta_{a,z}$ stands for the frequency that the decision maker takes an action when the available set of sources is $z$. Hence, it is obvious that 
\begin{equation}\label{eq:constraints_availability}
\sum_{a\in\aSet}\beta_{a,z} = \alpha_z\quad \forall\,z\in\zSet,
\end{equation}
where the shorthand notation $\alpha_z \eqDef P_Z(z)$ is introduced for convenience. 

We are ready to state the main result.
\begin{theorem}[The optimal error exponent region in the general setup]
\label{thm:softErrorExponent}
Suppose both Assumption~\ref{a:oneStepLLR} and \ref{a:discrimination_1} hold. Define
\begin{equation}\label{eq:err_expnt_region}
\textstyle
\zeta\eqDef\bigtimes\limits_{m=0}^{M-1}\,\lbp\bigcup\limits_{\bbeta^m\in\mathcal{C}}\zeta_m(\bbeta^m)\rbp
\end{equation}
where 
\begin{equation}\label{eq:constraint_set_general}
\mathcal{C} 
\eqDef \lbp \bbeta\in\mathbb{R}^{|\aSet||\zSet|}_+\,\mv\, 
\text{$\bbeta$ satisfies \eqref{eq:constraints_budget} and \eqref{eq:constraints_availability}}
\rbp,
\end{equation}
and for each $m\in\mSet$, $\zeta_m(\bbeta^m)$ is defined on the top of the next page.
\begin{figure*}[!t]
\normalsize
\begin{align*}
\zeta_m(\bbeta^m)
&\eqDef \lbp \big\{ e_{m\vert\theta} \,\big\vert\, \theta\in\mSet\setminus\{m\}\big\}\in\mathbb{R}_+^{M-1}\,\mv\, e_{m\vert\theta}\leq e^*_{m\vert\theta}(\bbeta^m),\,\forall\,\theta\in\mSet\setminus\{m\}\rbp,\\
e^*_{m\vert\theta}(\bbeta^m) 
&\eqDef\sum_{a\in\mathcal{A}}\sum_{z\in\mathcal{Z}}\beta^m_{a,z}\KLD{P^{a,z}_{m}}{P^{a,z}_{\theta}}\quad\forall\, m\in\mSet,\,\forall\,\theta\in\mSet\setminus\{m\}.
\end{align*}
\hrulefill
\vspace*{4pt}
\end{figure*}
Then, any $M(M-1)$-tuple $\{e_{m\vert\theta}\}$ inside $\zeta$ can be achieved by our proposed test with a corresponding adaptive selection policy. Conversely, an achievable $M(M-1)$-tuple $\{e_{m\vert\theta}\}$ must lie inside the region $\zeta$.
\end{theorem} 
\begin{IEEEproof}[Sketch of proof]
The proposed test and the achievability proof are provided in Section~\ref{sec:achievability}. 
The proof of the converse part of Theorem~\ref{thm:softErrorExponent} primarily relies on the data processing inequality and Doob's Optional Stopping Theorem, similar to that in \cite{LanWang_21}. Details can be found in Section~\ref{sec:converse}.
\end{IEEEproof}

Before moving on to the overview of our proposed scheme, let us make the following series of remarks on Theorem~\ref{thm:softErrorExponent}.

\begin{remark}
All the intuitions and discussions in Section~\ref{sec:pre_results} regarding Chernoff's setup remain valid in the general setup. For example, when $M=2$, there is no tradeoff between the error exponents, extending the result of \cite{LanWang_21}.
\end{remark}

\begin{remark}
Note that the tradeoffs among error exponents largely depend on the constraint set $\mathcal{C}$ defined in \eqref{eq:constraint_set_general}. From the expression, it can be seen that besides the heterogeneity among data sources, parameters of the problem formulation (the action space, the temporal availability, and the budget constraints) also influence the tradeoffs. For instance, if the parameters of the problem formulation allow the decision maker to select all data sources simultaneously at each time slot, there will be no tradeoffs among error exponents. 
\end{remark}

\begin{remark}[Convexity of the optimal error exponent region $\zeta$]\label{re:convex_of_zeta}
The convexity of $\zeta$ follows naturally from the direct product in \eqref{eq:err_expnt_region}  and the convexity of the sub-regions $\bigcup_{\bbeta^m\in\mathcal{C}}\zeta_m(\bbeta^m)$ for all $m\in\mSet$. The latter property can be shown by the definition of convexity, the fact that $\mathcal{C}$ is a convex set for any index set $\mathcal{I}$, and the fact that each $\zeta_m(\bbeta^m)$ is a hyper-rectangle with boundaries being linear functions of $\bbeta^m$, that is, $\big\{ e^*_{m|\theta}(\bbeta^m)\, \big\vert\, m\in\mSet, \theta\in\mSet\setminus\{m\}\big\}$.
\end{remark}

\begin{remark}[Computation of the optimal error exponent region $\zeta$]
The complexity of computing the optimal error exponent region $\zeta$ largely depends on the constraint set $\mathcal{C}$ as well. 
If $\mathcal{C}$ is complex, determining $\zeta$ might be computationally intractable. 
To illustrate, let us first consider the case where $|\mathcal{I}|<\infty$. In this setting, $\mathcal{C}$ becomes a convex polytope. 
Since each $\zeta_m(\bbeta^m)$ is a hyper-rectangle with boundaries being linear functions of $\bbeta^m$, 
the set $\bigcup_{\bbeta^m\in\mathcal{C}}\zeta_m(\bbeta^m)$ contains only a finite number of extreme points and is a convex polytope. Recall the definition of an extreme point: for any set $A$, a point $a\in A$ is an extreme point of the set $A$ if for any $b,c\in A$, $b\neq c$ and for any $\lambda\in(0,1)$, the point $a$ satisfies $a\neq\lambda b +(1-\lambda)c$.
Consequently, due to the direct product form of $\zeta$, it is also a convex polytope. This leads to the following equivalent characterization of $\zeta$.
\begin{proposition}[$\zeta$ is a convex polytope]\label{thm:polytope}
\[\textstyle
    \zeta=\bigtimes\limits_{m=0}^{M-1}\,\text{Conv}\lbp\text{extr}\lp \bigcup_{\bbeta^m\in V_{\mathcal{C}}}\zeta_m(\bbeta^m)\rp\rbp
\]
where $V_{\mathcal{C}}$ denotes the set of all extreme points of the constraint set $\mcal{C}$ when $\lvert\mathcal{I}\rvert<\infty$.
\end{proposition}
The proof of the proposition is given in Appendix~\ref{app:polytope}. One key implication of Proposition~\ref{thm:polytope} is that if $\mcal{C}$ has a simple structure (for example, when $\mcal{C}=\mathcal{P}_n$, the probability simplex), finding $V_{\mcal{C}}$ could be straightforward. In such cases, one can efficiently compute
\[\textstyle
\text{extr}\lp \bigcup_{\bbeta^m\in V_{\mathcal{C}}}\zeta_m(\bbeta^m)\rp
\]
and apply a off-the-shelf convex hull algorithm to compute $\zeta$. However, when $\mathcal{C}$ is more complex, constructing $V_{\mcal{C}}$ may become computationally intractable. As a result, computing $\zeta$ is also hard. 
\end{remark}

\subsection{Overview of the proposed test}\label{sec:overview}
For the achievability part of Theorem~\ref{thm:softErrorExponent}, from the expression in \eqref{eq:err_expnt_region}, the goal is to show that for any selection frequency tuples $\bbeta^0,\bbeta^1,\ldots,\bbeta^{M-1} \in \mathcal{C}$, the tuple of $M(M-1)$ individual error exponents $\Big\{ e_{m\vert\theta} \,\Big\vert\, m\in\mSet, \theta\in\mSet\setminus\{m\}\Big\}$ is achievable if $\big\{ e_{m\vert\theta} \,\big\vert\, \theta\in\mSet\setminus\{m\}\big\} \in \zeta_m(\bbeta^m)$ $\forall\,m\in\mSet$. 
Recall that $\mathcal{C}$ is the constraint set defined in \eqref{eq:constraint_set_general} in Theorem~\ref{thm:softErrorExponent} and each tuple $\bbeta^m$, $m\in\mSet$, comprises components indexed by the action $a$ and the available subset $z$, hinting the probability of taking action $a$ when the available subset is $z$ under a randomized policy.

As for the strategy of taking actions, note that $\bbeta^\theta$ (here we change the indexing convention from $m$ to $\theta$ for notational convenience) 
corresponds to the action taking policy under hypothesis $\theta$, and they are in general different across all $\theta\in\mSet$. 
Chernoff's strategy \cite{Chernoff_59} that \emph{exploits} the information accumulated so far is to first find a maximum likelihood estimate (MLE) $\hat{\theta}_{\text{MLE}}\in\mSet$ and employ the policy corresponding to $\bbeta^{\hat{\theta}_{\text{MLE}}}$. In our design, in addition to exploitation, we further mix it with a small (vanishing-in-$T$) probability of \emph{exploration} inspired by \cite{Vaidhiyan2015}, which is a common idea in the literature of sequential decision making problems such as bandits. This probability of exploration and the rate of \emph{uniformly} exploring all sources are carefully chosen to meet the constraints and achieve the target error exponents. 

As for the stopping rule and the inference strategy, it follows those in the multiple-hypothesis sequential probability ratio tests (MSPRT) \cite{DragalinTartakovsky_99} with thresholds designed judiciously to achieve the target error exponents and meet the constraints. A key statistic here is the cumulative log-likelihood ratio  
\[
S_{t,\theta,m}\eqDef\log\frac{\msf{P}_{\theta}^{(t)}(\info_t)}{\msf{P}_{m}^{(t)}(\info_t)},
\]
where $\msf{P}_{\theta}^{(t)}(\cdot)$ represents the probability distribution of $\info_t$, the accumulated information up to the end of time instance $t$ (see Definition~\ref{def:decision_maker}) under hypothesis $\theta$. 
The expression of $S_{t,\theta,m}$ will be further expanded in Appendix~\ref{app:pf_lem_S_is_martingale}.

To this end, let us sketch the proposed test, with more details left in Section~\ref{sec:details_of_proposed_test}. 
For a given $(\bbeta^0,\bbeta^1,\ldots,\bbeta^{M-1})$, at each time instance $t\geq1$, upon $Z_t$ being revealed, the decision maker carries out the following:
\begin{enumerate}
    \item Action taking policy and sample collection: 
    \par
    At the end of time slot $t-1$, find the MLE $\hat{\theta}_{\text{MLE}}$, that is, $\theta$ such that $\min_{m\neq\theta}S_{t-1,\theta,m}\geq0$. 
Based on this $\hat{\theta}_{\text{MLE}}$ and the corresponding $\bbeta^{\hat{\theta}_{\text{MLE}}}$ together with a judiciously designed exploration rate, we then set up a conditional PMF $P_{A_t|Z_t, \info_{t-1}}(\cdot| z, f_{t-1})$ for randomized action taking. 
How to design this conditional PMF is elaborated later. 
Subsequently, a random action $a$ is drawn from the conditional PMF, and the decision maker collects a tuple of samples $\mathbf{X}_t$ accordingly. 

    \item Stopping and inference rule: 
    \par
    If there exists a hypothesis $\theta$ such that $S_{t,\theta,m}$ exceeds a set of designed thresholds $\forall\,m\neq\theta$, then stop and declare such a $\theta$ as the inferred result.
\end{enumerate}

As for the conditional PMF used for action taking, if hypothesis $\theta$ has the largest cumulative log-likelihood ratio $(\hat{\theta}_{\text{MLE}}=\theta)$ at the end of time slot $t-1$, then we set
\begin{align}
\notag
&P_{A_t|Z_t, \info_{t-1}}(a| z, f_{t-1}) = \\
&\begin{cases}
(1-\epsilon(T))\frac{\beta^\theta_{a,z}}{\alpha_z}+\epsilon(T)\frac{1}{h\alpha_z}, 
&\text{if } a\in\mathcal{A}\setminus\{\emptyset\}\\[1ex]
1-\sum_{a'\in\mathcal{A}\setminus\{\emptyset\}}P_{A_t|Z_t, \info_{t-1}}(a'| z, f_{t-1}), 
&\text{if } a = \emptyset
\end{cases}
\label{eq:action_policy_rand}
\end{align}
where for notational convenience, $\alpha_z\eqDef P_Z(z)$, the probability of set $z$ being available, $\forall\,z\in\zSet$. The probability of conducting exploration $\epsilon(T)$ is judiciously designed to attain the asymptotic optimality of the test, and it vanishes as $T\ra\infty$. How to choose $\epsilon(T)$ will be elaborated in Section~\ref{sec:achievability}. 
The \emph{exploration rate} $\frac{1}{h}>0$ is chosen to be bounded as
\begin{equation}\label{eq:exploration_rate}
\frac{1}{h}\leq\min\lp\ \frac{\min_{z\in\zSet}\alpha_z}{\lvert\aSet\rvert}\ ,\ \inf_{i\in\mathcal{I},r_i>0}\frac{r_i}{b_i\lp\omega_1(\mathbf{1}),\ldots,\omega_n(\mathbf{1})\rp}\ \rp, 
\end{equation}
where each $r_i$ and $b_i(\cdot)$ are defined in Section~\ref{subsec:exponents}, each $\omega_j(\bbeta)$ is defined in \eqref{eq:select_freq_indv_src}, and $\mathbf{1}\eqDef\{1,\ldots,1\}\in\mathbb{R}^{|\aSet||\zSet|}_+$.
The upper bound of $\frac{1}{h}$ indicates that the exploration rate cannot exceed the probability of the minimum availability of the data sources and must satisfy the expected budget constraints. Note that in the upper bound of $1/h$ in \eqref{eq:exploration_rate}, the second term in the ``$\min$'' operation is taking the infimum over $i\in\mathcal{I}$ such that $r_i>0$. For those budget constraints indexed by $i\in\mathcal{I}$ with $r_i=0$, the decision maker can exclude actions involving data sources contributing to the cost computation in the linear function $b_i(\cdot)$. We can then design the conditional PMF for the action taking policy according to the effectively smaller action space. Since the effective action space becomes smaller and what we want is $1/h$ times the cardinality of the effective action space to be no more than $\alpha_z$ for all $z\in\zSet$, it is still a valid choice to take the denominator of the first term in the ``$\min$'' operation of the upper bound of $1/h$ in \eqref{eq:exploration_rate} to be the cardinality of the original action space. 

Before we move on to further technical details of the proposed scheme and the full proof of the achievability, let us close this section with some discussions about our proposed scheme. 
Note that our proposed scheme achieves optimality under Assumption~\ref{a:discrimination_1}, which is less stringent compared to the following assumption that was imposed in Chernoff’s original work \cite{Chernoff_59}. 
\begin{assumption}
\label{a:discrimination_strong}
    For any two hypotheses $m,\theta\in\mSet$, $\theta\neq m$, 
\[
    \forall\, z\in\zSet,\ \forall\, a\in\aSet \quad \KLD{P^{a,z}_{m}}{P^{a,z}_{\theta}}>0.
\]
\end{assumption}
It is interesting to note that under Assumption~\ref{a:discrimination_strong}, it suffices to employ ``pure exploitation,'' that is, $\epsilon(T)=0$, to achieve optimality. With that being said, technically speaking, exploration is introduced so that achievability can be proved under Assumption~\ref{a:discrimination_1}. 

Note that under Assumption~\ref{a:discrimination_1}, for any two distinct hypotheses, it is essential to develop techniques that enable the decision maker to select data sources capable of distinguishing between them, thereby ensuring vanishing individual error probabilities. 
In the literature, alternative methods have been proposed towards this goal. In \cite{NaghshvarJavidi_13}, a scheme with a warm-up phase is proposed, where actions are taken according to a predefined selection frequency before a specified time slot and then is adjusted to follow the maximum likelihood estimate selection afterwards. 
In \cite{NitinawaratAtia_13}, an alternative way for exploration is employed, where at preset time instances the decision maker uniformly select a source. 
We fail to extend these methods to prove the achievability in our problem setup. 
Our proposed method is similar to that in \cite{Vaidhiyan2015}. The major difference is that the probability of exploration in \cite{Vaidhiyan2015} is fixed, while our $\epsilon(T)$ vanishes with $T$ and the decaying rate is judiciously designed. As a result, the method in \cite{Vaidhiyan2015} is asymptotically sub-optimal, while our proposed method not only effectively controls budget constraints using an intuitive exploration probability but also provides a rigorous proof for achieving the asymptotic optimal error probabilities.

\section{Proof of Achievability}\label{sec:achievability}
In this section, the details of the proposed test are given, followed by the analysis of achievability.

\subsection{Details of the proposed test}\label{sec:details_of_proposed_test}
\subsubsection*{Notations}
The summation $\sum_{a,z}$ without its index set, is a short-hand notation for $\sum_{a\in\aSet\setminus\{\emptyset\}}\sum_{z\in\zSet}$. Note that we exclude the empty set because when $a=\emptyset$, $\KLD{P^{a,z}_{\theta}}{P^{a,z}_{m}}$ is defined as zero by convention for any $z\in\zSet$, $\theta,m\in\mSet$. The notations used in the proposed test are summarized in Table~\ref{table:Notations}.

{\renewcommand{\arraystretch}{2}
\begin{table}[ht!]
\centering
\begin{tabular}{|c| c|} 
 \hline
 Notation & Expression \\ [0.5ex] 
 \hline
 $\info_t$ &  $\lbp\lbp\mathbf{X}_s\rbp_{s=1}^{t}, \lbp A_s\rbp_{s=1}^{t}, \lbp Z_s\rbp_{s=1}^{t}\rbp$\\
 $S_{t,\theta,m}$ & $\log\frac{\msf{P}_{\theta}^{(t)}(\info_t)}{\msf{P}_{m}^{(t)}(\info_t)}$ \\
 $l$ & $\frac{1}{T^{1/6}}\lp\frac{\min_{\theta\in\mSet}\min_{m\neq\theta}A_{\theta,m}}{\min_{\theta\in\mSet}\min_{m\neq\theta}\Tilde{A}_{\theta,m}}\rp^2$ \\ 
 $\epsilon(T)$ & $\lp\log T\rp^{-\frac{1}{4}}$ \\
 $\Tilde{A}_{\theta,m}$ & $\frac{1}{h}\sum_{a,z}\KLD{P^{a,z}_{\theta}}{P^{a,z}_{m}}$ \\
 $A_{\theta,m}$ & $\frac{\lp1-\epsilon(T)\rp e^*_{\theta|m}\lp\bbeta^\theta\rp+\epsilon(T)\Tilde{A}_{\theta,m}}{1+l}$ \\
 $I(T)$ & $\epsilon(T)\min_{\theta\in\mSet}\min_{m\neq\theta}\Tilde{A}_{\theta,m}$ \\
 $b(l)$ & $\frac{l^3}{4(1+l)^3}\lp\frac{I(T)}{4 L}\rp^2$ \\
 $B(l)$ & $2+\frac{2(1+l)^2}{l^2}\lp\frac{4L}{I(T)}\rp^2$ \\
 $q(l,n,I(T))$ & $1+\frac{1}{b(l)}\Big(1+\log\lp MB(l)\lp1+b(l)\rp\rp\Big)$ \\
 $\Delta$ & $-MB(l)\lp1+b(l)\rp e^{-b(l)(T-1)}$\\
  $K(x)$ & $\sqrt{xe+1}-1$ with domain:$[-e^{-1},0]$ and range:$[-1,0]$\\ 
 $\epsilon_\theta$ & $A^{\text{max}}_{\theta}\lp1-\frac{K\lp\Delta\rp}{b(l)}\rp$ \\
  $A^{\text{max}}_{\theta}$ & $\max_{m\neq\theta}A_{\theta,m}$\\[1ex] 
 \hline
\end{tabular}
\caption{Summary of notations in the proposed test.}
\label{table:Notations}
\end{table}
}

Let $e$ be Euler's number ($e\cong2.71828...$), $q\lp l,n,I(T)\rp$ be defined in Table~\ref{table:Notations}, $n$ be the total number of data sources, $l$ and $I(T)$ be two specific sequences (both indexed by $T$) defined in Table~\ref{table:Notations}. Depending on the expected stopping time constraint $T$, our proposed test comprises two regimes:
\begin{itemize}
    \item Regime 1: $T<\max\{e, q\lp l,n,I(T)\rp\}$. The decision maker stops at time instance $t=1$, does not select any sources, and randomly guesses a hypothesis.
    \item Regime 2: $T\geq\max\{e, q\lp l,n,I(T)\rp\}$. At each time instance $t\geq1$, the decision maker carry out the test proposed below.
\end{itemize}
We distinguish the scheme into two different regimes so that the expected stopping time and budget constraints can be satisfied for all values of $T$, not just asymptotically. 
One may concern that the random guess in Regime 1 affects the error exponent. Nevertheless, since we only consider the asymptotic error performance (when $T\to\infty$) and $q\lp l,n,I(T)\rp$ is sub-linear in $T$ (see Lemma~\ref{le:f_is_subl} later), the random guess does not dominate the asymptotes of error probabilities. Details are given in Section~\ref{sec:analysis_of_proposed_test}.

Let us now provide the detailed procedures of the test in the second regime (overviewed in Section~\ref{sec:overview}) as follows. Given any $\bbeta^0,\bbeta^1,\ldots,\bbeta^{M-1} \in \mathcal{C}$ (defined in \eqref{eq:constraint_set_general}), at each time instance $t\geq1$:
\begin{enumerate}
\item
Firstly, the decision maker conducts the randomized action taking following the PMF $P_{A_t|Z_t, \info_{t-1}}(\cdot| z, f_{t-1})$ with the MLE $\hat{\theta}_{\text{MLE}}$ at time slot $t-1$, under a revealed $Z_t=z\in\zSet$. Specifically, if $\hat{\theta}_{\text{MLE}}=\theta$ at time slot $t-1$, we design $P_{A_t|Z_t, \info_{t-1}}(a| z, f_{t-1})$ as in \eqref{eq:action_policy_rand} with the \emph{exploration rate} $\frac{1}{h}>0$ satisfies \eqref{eq:exploration_rate}. 
Then, the decision maker collects a tuple of samples drawn from the selected data sources, namely, 
\[
\mathbf{X}_t\eqDef\{X_{t,j}\,\vert\, j\in A_t\cap Z_t\}.
\]

\item
Secondly, after collecting $\mathbf{X}_t$, the decision maker decides to stop when there exists $\theta\in\mSet$ such that $\forall\,m\in \mSet\setminus\{\theta\}$, the cumulative log likelihood ratio
\begin{equation}\label{eq:stopping_criterion}
S_{t,\theta,m} \ge \lp TA_{\theta,m}-\epsilon_\theta\rp,
\end{equation} 
and declares that the inferred hypothesis is the $\theta$ that satisfies \eqref{eq:stopping_criterion}. Note that $S_{t,\theta,m}$, $A_{\theta,m}$ and $\epsilon_\theta$, the parameters related to thresholding the log likelihood ratio, are defined in Table~\ref{table:Notations}. 
Otherwise, it will proceed to the next time slot. In words, the stopping time 
\[
\begin{aligned}
&\tau \eqDef\\
&\inf\Bigg\{ t\geq1\,\Bigg\vert\, \begin{array}{l}
\exists\,\theta\in\mSet\ \text{such that}\\
\min\limits_{m\neq\theta}\lbp S_{t,\theta,m}-\lp TA_{\theta,m}-\epsilon_\theta\rp\rbp\geq0
\end{array}\Bigg\}.
\end{aligned}
\]
\end{enumerate}

Before we proceed to the detailed analysis, let us remark that the design of the proposed scheme is very natural. 
For the stopping and inference rules, we follow the concept of MSPRT \cite{DragalinTartakovsky_99} with thresholding based on the value to which with high probability the cumulative log likelihood ratio converges as $t$ gets close to the sufficiently large $T$. 
This thresholding also guarantees that the desired error exponents can be attained as observed in passive sequential hypothesis testing \cite{WaldWolfowitz_48,DragalinTartakovsky_99},  \cite[Chapter 16.3]{polyanskiy2025information}. 
Regarding the action taking policy, apart from the designed exploration mechanism, we adhere to the maximum likelihood estimate selection method introduced in Chernoff’s work \cite{Chernoff_59}. As discussed in the end of Section~\ref{sec:overview}, the main novelty lies in the exploration, and it leads to optimal achievability under a less stringent assumption.

\subsection{Analysis of the proposed test}\label{sec:analysis_of_proposed_test}

The proof of achievability comprises three parts: 
\begin{enumerate}
\item 
Demonstrating that our proposed test satisfies the expected stopping time constraints with the non-asymptotic analysis slightly modified from the achievability proofs of \cite[Proposition 2]{NaghshvarJavidi_13} and \cite[Theorem 4]{Vaidhiyan2015}, which originated from \cite[Theorem 1]{Chernoff_59}. 
\item 
Showing that our proposed test satisfies the expected budget constraints, that is, constraint \eqref{eq:linear_exp_budg_constr} is met. 
\item 
Verifying that, with a given $\{\bbeta^i\,\vert\,\bbeta^i\in\mathcal{C},\,i\in\mSet\}$, our proposed test achieves the tuples of $M(M-1)$ individual error exponents $\big\{ e_{m\vert\theta} \,\big\vert\, m\in\mSet, \theta\in\mSet\setminus\{m\}\big\}$ 
satisfying 
\[
\big\{ e_{m\vert\theta} \,\big\vert\, \theta\in\mSet\setminus\{m\}\big\} \in \zeta_m(\bbeta^m)
\]
for every $m\in\mSet$.
\end{enumerate}
Note that Part 2) and 3) are similar to the achievability proof of \cite[Theorem 1]{LanWang_21}.

Throughout the proof of achievability, we are given a set of $\bbeta^0,\bbeta^1,\ldots,\bbeta^{M-1} \in \mathcal{C}$. Let us detail the three parts separately in the following.

\subsubsection{Expected stopping time constraints}\label{sec:exp_time_constraint}
In this part, the goal is to show that for $T\geq 1$,
\begin{equation}\label{eq:pf_achieve_part1}
    \E_\theta[\tau]\leq T\quad\forall\,\theta\in\mSet.
\end{equation} 
It is straightforward that the above is satisfied in Regime 1 of the proposed test. Hence in the following, we focus on Regime 2, that is, $T\geq\max\{e, q\lp l,n,I(T){}\rp\}$.

Let us first define a special stopping time $\tau_\theta$ for any $\theta\in\mSet$ as follows:
\[
    \tau_\theta\eqDef\inf\lbp t\,:\,\min_{m\neq\theta}\lbp S_{t,\theta,m}-TA_{\theta,m}+\epsilon_\theta\rbp\geq 0\rbp
\]
Therefore, by the definition of the stopping time $\tau$ in our proposed test, we have $\tau=\min_{\theta\in\mSet}\tau_\theta$, 
leading to $\E_{\theta}\lb\tau\rb\le \E_{\theta}\lb\tau_\theta\rb$, $\forall\,\theta\in\mSet$. 
Hence, it suffices to show that in Regime 2 of our proposed test,
\begin{equation}\label{eq:pf_achieve_part1_suff}
    \E_\theta\lb\tau_\theta\rb\leq T\quad \forall\,\theta\in\mSet.
\end{equation}

To proceed, for each $\theta\in\mSet$, we decompose $\E_\theta\lb\tau_\theta\rb$ into two terms and upper bound them as 
\[
    \E_\theta\lb\tau_\theta\rb\le\kappa_{\theta,T,l}+\sum_{k=\kappa_{\theta,T,l}}^{\infty}\Pr_\theta\lp\tau_\theta > k\rp,
\]
where 
\[
\kappa_{\theta,T,l}\eqDef \left\lceil T-\frac{\epsilon_\theta}{A^{\text{max}}_{\theta}}\right\rceil,\quad
A^{\text{max}}_{\theta}\eqDef\max_{m\neq\theta}A_{\theta,m},
\]
and $\epsilon_\theta$ and $A_{\theta,m}$ are defined in Table~\ref{table:Notations}.
In the following, we are going to first give a key lemma stating that the probability $\Pr_\theta\lp\tau_\theta > k\rp$ decays exponentially as $k\rightarrow\infty$. 
With the lemma, it is then simple to show that 
$\sum_{k=\kappa_{\theta,T,l}}^{\infty}\Pr_\theta\lp\tau_\theta > k\rp \le \frac{\epsilon_\theta}{A^{\text{max}}_{\theta}}-1$ if $T\geq\max\{e, q\lp l,n,I(T)\rp\}$, leading to the conclusion that \eqref{eq:pf_achieve_part1_suff} is met, and consequently \eqref{eq:pf_achieve_part1}.

\begin{lemma}\label{thm:lemma1} 
For any $k\ge\kappa_{\theta,T,l}$ and $\theta\in\mSet$,
\[
\Pr_\theta\lp\tau_\theta > k\rp \le M B(l) e^{-kb(l)}
\]
where $l$, $b(l)$, and $B(l)$ are defined in Table~\ref{table:Notations}. 
\end{lemma}
The proof of Lemma~\ref{thm:lemma1} is given in Section~\ref{sec:proof_of_lemma1}.

By Lemma~\ref{thm:lemma1} and the fact that $b(l)$ and $B(l)$ are positive, we upper bound $\sum_{k=\kappa_{\theta,T,l}}^{\infty}\Pr_\theta\lp\tau_\theta > k\rp$ as follows:
\begin{align}
\notag	&\sum_{k=\kappa_{\theta,T,l}}^{\infty}\Pr_\theta\lp\tau_\theta > k\rp\\ 
\notag	&\le MB(l)\frac{e^{-\kappa_{\theta,T,l}b(l)}}{1-e^{-b(l)}}\\
\notag	&\le MB(l)\lp 1+\frac{1}{b(l)} \rp e^{-\kappa_{\theta,T,l}b(l)}\\
\notag	&= MB(l)\lp 1+\frac{1}{b(l)} \rp e^{-\left\lceil T-\frac{\epsilon_\theta}{A^\text{max}_\theta}\right\rceil b(l)}\\
\notag    	&= MB(l)\lp 1+\frac{1}{b(l)} \rp e^{-\left\lceil T-\lp 1-\frac{K(\Delta)}{b(l)} \rp\right\rceil b(l)}\\
\notag    	&\le MB(l)\lp 1+\frac{1}{b(l)} \rp e^{-\lp T-\lp 1-\frac{K(\Delta)}{b(l)} \rp\rp b(l)}\\
\notag   	&= \frac{1}{b(l)}MB(l)\lp b(l)+1 \rp e^{-b(l)\lp T-1 \rp}e^{-K(\Delta)}\\
\notag    	&=-\frac{1}{b(l)}\Delta e^{-K(\Delta)}\\
	&=-\frac{1}{b(l)}W_0(\Delta)e^{W_0(\Delta)}e^{-K(\Delta)}\label{eq:extra_10}
\end{align}
where $W_0(\cdot)$ is the principal branch of the Lambert function. Recall that $\Delta\eqDef-MB(l)\lp b(l)+1 \rp e^{-b(l)\lp T-1 \rp}$ as defined in Table~\ref{table:Notations}. Also note that for $T\geq\max\{e, q\lp l,n,I(T)\rp\}$, $\Delta$ satisfies $-e^{-1}\leq\Delta<0$ and hence falls in the domain of $W_0(\cdot)$.

Finally, we show that $\eqref{eq:extra_10}\leq\frac{\epsilon_\theta}{A^{\text{max}}_{\theta}}-1=-\frac{K(\Delta)}{b(l)}$ to complete the proof. By \cite[Theorem 3.2]{Roig2020},
\[
    K(x)\le W_0(x)\quad \text{for}\ -e^{-1}\le x< 0,
\]
and hence we have $W_0(\Delta)\ge K(\Delta)\ge-1$. Due to the fact $xe^x$ is increasing when $x\ge-1$, it is straightforward that $W_0(\Delta)e^{W_0(\Delta)}\ge K(\Delta)e^{K(\Delta)}$, completing the proof of the first part.

\subsubsection{Linear expected budget constraints}\label{sec:budg_cons}
Our goal is to show that the proposed test satisfies the expected budget constraints, 
that is
\[
b_i\lp B_1^\theta, B_2^\theta, \ldots, B_n^\theta \rp \le U_i \equiv r_i T\quad\forall\,i\in\mcal{I}, \forall\,\theta\in\mSet,
\]
where $B^\theta_j \eqDef \E_\theta\lb\sum_{t=1}^{\tau}\Indc{j\in A_t\cap Z_t}\rb$ for each $\theta\in\mSet$ and $j\in\wSet$, and $b_i(\cdot)$, $U_i$, and $r_i$ for each $i\in\mathcal{I}$ are defined in Section~\ref{sec:formulation}.

Let us first introduce a stochastic process $\lbp M_{t,i} \,\mv\, t\ge 0\rbp$ for the linear constraint function $b_i(\cdot)$ defined in Section~\ref{sec:formulation} as follows: $\forall\,i\in\mathcal{I}$,
\[
    M_{t,i}\eqDef b_i\lp B_{t,1}, B_{t,2}, \ldots, B_{t,n} \rp-r_i t\quad\forall\,t\ge0,
\]
where each $B_{t,j}\eqDef \sum_{s=1}^{t}\Indc{j\in A_t\cap Z_t}$ denotes the number of times that source $j\in\wSet$ is selected up to time instance $t\ge 1$ and $M_{0,i}=0$. Hence, we can note that $\E_\theta\lb B_{\tau, j}\rb = B^\theta_j$ for each $\theta\in\mSet$ and $j\in\wSet$. 

We then show that this stochastic process is a super martingale adapted to the natural filtration $\{\mathcal{F}_t\eqDef\sigma(\info_{t})\}_{t\ge0}$ under any given hypothesis $\theta\in\mSet$. Specifically, by the linearity of $b_i(\cdot)$, we have
\begin{align}
\notag    &\E_\theta\lb M_{t+1,i}\,\mv\,\mathcal{F}_t\rb\\ 
\notag    &= \E_\theta\lb b_i\lp B_{t+1,1}, B_{t+1,2}, \ldots, B_{t+1,n} \rp - \lp t+1\rp r_i\,\mv\,\mcal{F}_t \rb\\
\notag    &=b_i(B_{t,1}, B_{t,2}, \ldots, B_{t,n})-r_it\\
\notag    &\quad + \E_\theta\lb b_i\lp \lbp \Indc{j\in A_{t+1}\cap Z_{t+1}} \rbp_{j=1}^{n} \rp\,\mv\, \mcal{F}_t \rb- r_i \\
\notag    &=M_{t,i} + \E_\theta\lb b_i\lp \lbp \Indc{j\in A_{t+1}\cap Z_{t+1}} \rbp_{j=1}^{n} \rp\,\mv\, \mcal{F}_t \rb- r_i \\
    &\leq  M_{t,i} \label{eq:per_const_1}
\end{align}
where \eqref{eq:per_const_1} holds true by the linearity of $b_i(\cdot)$, 
\begin{align}
\notag    &\E_\theta\lb b_i\lp \lbp \Indc{j\in A_{t+1}\cap Z_{t+1}} \rbp_{j=1}^{n} \rp\,\mv\,\mcal{F}_t \rb\\
\notag    &=b_i\lp \lbp \E_\theta\lb \Indc{j\in A_{t+1}\cap Z_{t+1}}\,\mv\,\mcal{F}_t \rb \rbp_{j=1}^{n} \rp\\
\notag    &=b_i\lp\lbp\sum_{a,z}P_{A_{t+1}|Z_{t+1}, \info_{t}}\lp a\vert z, f_t\rp\alpha_z\Indc{j\in a\cap z}\rbp_{j=1}^{n}\rp\\
    &\le\max_{\theta\in\mSet}b_i\lp\lbp\sum_{a,z}\lp\begin{subarray}{l}\lp 1-\epsilon\lp T\rp \rp\beta^\theta_{a,z}\\+\epsilon(T)/h\end{subarray}\rp\Indc{j\in a\cap z}\rbp_{j=1}^{n}\rp\label{eq:per_const_2}\\
\notag    &=\max_{\theta\in\mSet} \lp 1- \epsilon(T) \rp b_i\lp\omega_1(\bbeta^\theta), \omega_2(\bbeta^\theta),\ldots,\omega_n(\bbeta^\theta)\rp\\
\notag    &\quad + \epsilon(T)\frac{1}{h}b_i\lp \omega_1(\mathbf{1}), \omega_2(\mathbf{1}), \ldots, \omega_n(\mathbf{1}) \rp\\
    &\le(1-\epsilon(T))r_i + \epsilon(T)r_i \label{eq:per_const_3}\\
\notag    &= r_i, 
\end{align}
where each $\omega_j(\bbeta)\eqDef\sum_{a\in\aSet}\sum_{z\in\zSet}\beta_{a,z}\Indc{j\in a\cap z}$ in Theorem~\ref{thm:softErrorExponent}.
Note that \eqref{eq:per_const_2} is due to the action taking policy defined in our proposed test, and \eqref{eq:per_const_3} is due to $\bbeta^\theta\in\mathcal{C}$ for all $\theta\in\mSet$ and the design of $\frac{1}{h}$ in our proposed test. Recall that $\boldsymbol{\omega}(\cdot)$ is defined in Theorem~\ref{thm:softErrorExponent}.

Consequently, by the fact that $\E_\theta[\tau]\le T$ as shown Part 1), the linearity of $b_i(\cdot)$, and the Optional Stopping Theorem, we have
\begin{align*}
0&=\E_\theta\lb M_{0,j}\rb
\ge\E_\theta\lb M_{\tau,j}\rb\\
&\ge\E_\theta\lb b_i\lp B_{\tau,1}, B_{\tau,2}, \ldots, B_{\tau,n} \rp\rb-r_iT\\
&= b_i\lp \lbp \E_\theta\lb B_{\tau,j}\rb \rbp_{j=1}^{n} \rp - r_iT.
\end{align*}
Plugging the definition of $B_{t,j}$, we have
\[
b_i\lp B_1^\theta,\ldots,B_n^\theta \rp \le r_i T\equiv U_i\quad \forall\,i\in\mcal{I}, \forall\,\theta\in\mSet, 
\]
completing the proof of this part.

\subsubsection{Achievable error exponents}
Recall that in our proposed test, there are two regimes depending on the value of $T$: $T<\max\{e, q\lp l,n,I(T)\rp\}$ (Regime 1) and $T\ge \max\{e, q\lp l,n,I(T)\rp\}$ (Regime 2). As shown in the following lemma, $q\lp l,n,I(T)\rp \in o(T)$, and hence for $T$ sufficiently large, the test operates in the second regime. As a result, for deriving the achievable error exponent in our asymptotic notion, we only need to consider the second regime.

\begin{lemma}\label{le:f_is_subl}
$q(l,n,I(T))$ is sublinear in $T$, that is, $q(l,n,I(T)) \in o(T)$
\end{lemma}
\begin{IEEEproof}
Recall the definition of $q(l,n,I(T))$ as well as the relevant terms in Table~\ref{table:Notations}:
\begin{align*}
q(l,n,I(T)) &\eqDef 1+\frac{1}{b(l)}\Big(1+\log\lp MB(l)\lp1+b(l)\rp\rp\Big),\\
l &\eqDef \frac{1}{T^{1/6}}\lp\frac{\min_{\theta\in\mSet}\min_{m\neq\theta}A_{\theta,m}}{\min_{\theta\in\mSet}\min_{m\neq\theta}\Tilde{A}_{\theta,m}}\rp^2\\
b(l) &\eqDef \frac{l^3}{4(1+l)^3}\lp\frac{I(T)}{4 L}\rp^2,\\
B(l) &\eqDef 2+\frac{2(1+l)^2}{l^2}\lp\frac{4L}{I(T)}\rp^2,\\
I(T) &\eqDef \epsilon(T)\min_{\theta\in\mSet}\min_{m\neq\theta}\Tilde{A}_{\theta,m},\\
\epsilon(T) &\eqDef \lp\log T\rp^{-\frac{1}{4}}.
\end{align*}
It suffices to show that $\frac{1}{b(l)}\log\lp MB(l)\lp1+b(l)\rp\rp\in o(T)$.
Note the following:
\begin{enumerate}
    \item $\Tilde{A}_{\theta,m}$, $A_{\theta,m}$ are finite for any two different hypotheses $\theta,m\in\mSet$ due to Assumption~\ref{a:oneStepLLR}.
    \item $M$, $L$ are also finite.
\end{enumerate}
By plugging the definitions of the respective terms, including $l$, $b(l)$, $B(l)$, $I(T)$, and $\epsilon(T)$, straightforward calculations lead to 
\[
\frac{1}{b(l)}\log\lp MB(l)\lp 1+b(l)\rp\rp\in O\lp T^{\frac{5}{6}}\lp\log(T)\rp^{\frac{1}{2}}\rp \subseteq o(T),
\]
and the proof is complete.
\end{IEEEproof}

Hence, given any hypotheses $\theta$, $\forall\,m\neq\theta$, when $T$ is sufficiently large, we have
\begin{align*}
    \pi_{m\vert\theta}
    &= \Pr_{\theta}\lp\min_{i\neq m}\lbp S_{\tau,m,i}-TA_{m,i}+\epsilon_m\rbp\ge0\rp \\
    &\le \Pr_{\theta}\lp S_{\tau,m,\theta}-TA_{m,\theta}+\epsilon_m\ge0\rp \\
    &= \Pr_{\theta}\lp e^{S_{\tau,m,\theta}} \ge e^{TA_{m,\theta}-\epsilon_m}\rp \\
    &\le \frac{\E_\theta\lb e^{S_{\tau,m,\theta}}\rb}{e^{TA_{m,\theta}-\epsilon_m}}.
\end{align*}

To proceed, it is shown in the next lemma that $\lbp e^{S_{t,m,\theta}}\,\mv\, t\geq0\rbp$ is a martingale.

\begin{lemma}\label{le:S_is_martingale}
The stochastic process $\lbp e^{S_{t,m,\theta}}\,\mv\, t\geq0 \rbp$ is a martingale adapted to the natural filtration $\lbp \mathcal{F}_{t}\rbp_{t\geq0}$ under hypothesis $\theta$, $\forall\,\theta,m\in\mSet$, $\theta\neq m$. 
\end{lemma}
The proof of Lemma~\ref{le:S_is_martingale} is given in Appendix~\ref{app:pf_lem_S_is_martingale}.

Finally, by the Optional Stopping Theorem, $\E_\theta\lb e^{S_{\tau,m,\theta}} \rb=\E_\theta\lb e^{S_{0,m,\theta}}\rb=1$, and hence
\[
\pi_{m\vert\theta} \le e^{-T\lp A_{m,\theta}-\frac{\epsilon_m}{T}\rp}
\]
for $T$ sufficiently large. 
Therefore, $\forall\,\theta,m\in\mSet$, $\theta\neq m$, we can conclude that
\[
    \liminf_{T\to\infty}\lbp\frac{1}{T}\log\frac{1}{\pi_{m\vert\theta}}\rbp\ge \liminf_{T\to\infty}\lbp A_{m,\theta}-\frac{\epsilon_m}{T}\rbp=e^*_{m\vert\theta}(\bbeta^m)
\]
where $\epsilon_m\in o(T)$ as defined in Table~\ref{table:Notations} for any $m\in\mSet$. The proof of this part is complete.

\section{Proof of Converse}\label{sec:converse}
Let us first denote the binary KL divergence and the binary entropy function as follows: for $p,q\in [0,1]$,
\begin{align*}
\kld{p}{q} &\textstyle\eqDef p\log \frac{p}{q} + (1-p)\log\frac{1-p}{1-q} = \KLD{\Ber(p)}{\Ber(q)},\\
h(p) &\textstyle\eqDef p\log\frac{1}{p} + (1-p)\log\frac{1}{1-p} = H(\Ber(p)).
\end{align*}

To prove the converse part of Theorem~\ref{thm:softErrorExponent}, it suffices to show that, for any achievable $M(M-1)$ types of error exponents $\Big\{ e_{m\vert\theta} \,\Big\vert\, m\in\mSet, \theta\in\mSet\setminus\{m\}\Big\}$, there exist $\bbeta^0, \bbeta^1,...,\bbeta^{M-1}\in\mathcal{C}$ such that $\forall\, m\in\mSet$ $\forall\, \theta\in\mSet\setminus\{m\}$,
\[
     e_{m\vert\theta}\leq\sum_{a\in\aSet}\sum_{z\in\zSet}\beta^m_{a,z}\KLD{P^{a,z}_{m}}{P^{a,z}_{\theta}},
\]
thereby the $\Big\{ e_{m\vert\theta} \,\Big\vert\, m\in\mSet, \theta\in\mSet\setminus\{m\}\Big\}\in\zeta$, the optimal error exponent region defined in Theorem~\ref{thm:softErrorExponent}. 

The proof of converse is outlined as follows. First, we utilize the data processing inequality and Optional Stopping Theorem to find the upper bound of the $\frac{1}{T}\log\frac{1}{\pi_{m|\theta}}$ for any $m\in\mSet$, $\theta\in\mSet\setminus\{m\}$, $\forall\,T\ge1$, where the action taking \emph{frequency} that appears in the upper bound plays a key role. Then, we derive the necessary conditions for the action taking frequency to meet the linear expected budget constraints including expected stopping time constraints. We then observe that the collection of action taking frequencies satisfying the necessary conditions is the constraint set $\mathcal{C}$ defined in Theorem~\ref{thm:softErrorExponent}. Finally, we leverage the fact that $\mathcal{C}$ is a compact set to complete the proof.

Let us start the proof with an useful decomposition of $\log\frac{1}{\pi_{m|\theta}}$: $\forall\,m\in\mSet,\ \forall\,\theta\in\mSet\setminus\{m\}$,
\begin{equation}\label{eq:converse_step1}
\log\frac{1}{\pi_{m|\theta}}
=\frac{1}{\pi_{m|m}}\lp\!\!
\begin{array}{l}
\kld{\pi_{m|m}}{\pi_{m|\theta}}+h(\pi_{m|m})\\
+\Pr_m(\phi\neq m)\log\Pr_\theta(\phi\neq m)
\end{array}\!\!
\rp.
\end{equation}
The derivation of the above decomposition goes as follows:
\begin{align*}
\kld{\pi_{m|m}}{\pi_{m|\theta}}
&=\Pr_{m}(\phi=m)\log{\lp\frac{\Pr_{m}(\phi=m)}{\Pr_{\theta}(\phi=m)}\rp}\\
&\quad + \Pr_{m}(\phi\neq m)\log{\lp\frac{\Pr_{m}(\phi\neq m)}{\Pr_{\theta}(\phi\neq m)}\rp}\\
&= - h\lp \pi_{m\vert m} \rp - \pi_{m\vert m}\log\pi_{m\vert\theta}\\
&\quad -\Pr_m(\phi\neq m)\log\Pr_\theta(\phi\neq m).
\end{align*}
Rearranging terms leads to \eqref{eq:converse_step1}.

Next, by the data processing inequality, 
\begin{equation}
\label{eq:converse_step2}
\kld{\pi_{m|m}}{\pi_{m|\theta}}
\le \KLD{\msf{P}_{m}^{(\tau)}}{\msf{P}_{\theta}^{(\tau)}} = \E_m \lb S_{\tau,m, \theta} \rb,
\end{equation}
where
\begin{align*}
    S_{\tau,m, \theta}&= \sum_{s=1}^{\tau}\log\frac{P^{A_s,Z_s}_{m}(\mathbf{X}_{s})}{P^{A_s,Z_s}_{\theta}(\mathbf{X}_{s})}\\
    &= \sum_{s=1}^{\tau}\sum_{a\in\aSet}\sum_{z\in\zSet}\mathbbm{1}\{A_s=a,Z_s=z\}\log\frac{P^{a,z}_{m}(\mathbf{X}_{s})}{P^{a,z}_{\theta}(\mathbf{X}_{s})}.
\end{align*}
Note that this explicit form of the cumulative log-likelihood ratio was derived in Lemma~\ref{le:S_is_martingale}. 
Then, by Optional Stopping Theorem, we have
\begin{align}
\notag &\E_m \lb S_{\tau,m, \theta} \rb\\
&= \sum_{a\in\aSet}\sum_{z\in\zSet}\E_m \lb \sum_{s=1}^{\tau}\Indc{A_s=a,Z_s=z}\rb\KLD{P^{a,z}_{m}}{P^{a,z}_{\theta}}. \label{eq:converse_step3}
\end{align}

To this end, let us define the \emph{frequency} for any $a\in\aSet$ and $z\in\zSet$ as follows:
\begin{equation}
    \beta^{m,(T)}_{a, z} \eqDef \frac{\E_m \lb \sum_{s=1}^{\tau}\Indc{A_s=a,Z_s=z}\rb}{T}\quad \forall\,T\geq1 \label{eq:beta_def}.
\end{equation}
Let $\bbeta^{m,(T)}\eqDef\lbp\beta^{m, (T)}_{a,z}\,\mv\, a\in\aSet,z\in\zSet\rbp$. 
In the following, we show that 
\begin{equation}
\bbeta^{m,(T)}\in\mathcal{C}\quad \forall\, T\geq1,\, \forall\, m\in\mSet \label{eq:necessary_cond}
\end{equation}
are a set of necessary conditions that $\bbeta^{m,(T)},\ m\in\mSet$ should satisfy for the constraints in \eqref{eq:linear_exp_budg_constr} and \eqref{eq:expected_stopping_time_cons} to be met. Recall that $\mathcal{C}$ is defined in Theorem~\ref{thm:softErrorExponent}.
\begin{enumerate}
\item 
As for the linear expected budget constraints, plugging the definition of $B^m_j$ (see Section~\ref{subsec:exponents}), we have $\forall\,j\in\wSet$,
\begin{align*}
    &B^m_j\\
    &=\sum_{t=1}^{\infty}\E_m\lb\Indc{j\in A_t\cap Z_t}\Indc{t\leq\tau}\rb\\
    &=\sum_{t=1}^{\infty}\sum_{a,z}\lp\!\!\begin{array}{l}
    \Pr_m\lp A_t=a,Z_t=z\rp\Indc{j\in a\cap z}\\
    \E_m\lb\Indc{t\leq\tau}\,\mv\,A_t=a,Z_t=z\rb
    \end{array}\!\!\rp\\
    &=\sum_{a,z}\Indc{j\in a\cap z}\sum_{t=1}^{\infty}\Pr_m\lp A_t=a,Z_t=z,t\leq\tau\rp\\
    &=\sum_{a,z}\Indc{j\in a\cap z}\E_m\lb\sum_{t=1}^{\tau}\Indc{A_t=a,Z_t=z}\rb.
\end{align*}
Since $b_i(B^m_1,\ldots,B^m_n)\le U_i\ \forall\,i\in\mcal{I}$, and by the definition of $\bbeta^{m,(T)}$ in \eqref{eq:beta_def}, we have $\forall\,i\in\mcal{I}$, 
\[
b_i\lp\omega_1\lp\bbeta^{m,(T)}\rp,\ldots,\omega_n\lp\bbeta^{m,(T)}\rp\rp \leq \frac{U_i}{T}=r_i,
\]
which is obtained by dividing both sides of the inequality $b_i(B^m_1,\ldots,B^m_n)\le U_i\ \forall\,i\in\mcal{I}$ by $T$. Recall that each $\omega_j(\cdot)$ is defined in Theorem~\ref{thm:softErrorExponent} and each $b_i(\cdot)$ is a linear function defined in Section~\ref{subsec:exponents}.

\item
For the expected stopping time constraint, since $\E_m\lb\tau\rb\le T$, we have: 
$\forall\,z\in\zSet$,
\[\begin{aligned}
\sum_{a\in\aSet}\beta^{m,(T)}_{a, z}
&=\frac{\E_m\lb\sum_{s=1}^{\tau}\Indc{Z_s=z}\rb}{T}\\
&=\frac{\E_m\lb\tau\rb\alpha_z}{T}\leq\alpha_z.
\end{aligned}
\]
\end{enumerate}

From 1) and 2), we conclude that to satisfy both the linear expected budget constraints and the expected stopping time constraints, a necessary condition for $\bbeta^{m,(T)}$ is that $\bbeta^{m,(T)}\in\mathcal{C}\ \forall\,m\in\mSet, T\ge1$.

With the necessary conditions \eqref{eq:necessary_cond}, we combine them along with \eqref{eq:converse_step1}, \eqref{eq:converse_step2}, and \eqref{eq:converse_step3} to arrive at: $\forall T\geq 1$, $\forall\,m\in\mSet$, $\forall\,\theta\in\mSet\setminus\{m\}$,
\begin{align}
\notag &\frac{1}{T}\log\frac{1}{\pi_{m|\theta}}\\
\notag &\le\frac{1}{\pi_{m|m}}\lp
\begin{array}{l}
\sum_{a\in\aSet}\sum_{z\in\zSet} \beta^{m,(T)}_{a,z} \KLD{P^{a,z}_{m}}{P^{a,z}_{\theta}}\\
+ \frac{h\lp\pi_{m|m}\rp}{T} +\frac{\Pr_m(\phi\neq m)\log\Pr_\theta(\phi\neq m)}{T}
\end{array}
\rp\\
&\le\frac{1}{\pi_{m|m}}\lp\sum_{a\in\aSet}\sum_{z\in\zSet} \beta^{m,(T)}_{a,z} \KLD{P^{a,z}_{m}}{P^{a,z}_{\theta}}+\frac{h\lp\pi_{m|m}\rp}{T}\rp, \label{eq:converse_step4}
\end{align}
where \eqref{eq:converse_step4} holds since $\Pr_\theta(\phi\neq m)\leq 1$. 
We then take liminf on both sides of the inequality \eqref{eq:converse_step4} and get
\begin{align*}
&\liminf_{T\to\infty}\lbp\frac{1}{T}\log\frac{1}{\pi_{m|\theta}}\rbp\\
&\leq\liminf_{T\to\infty}\lbp\sum_{a\in\aSet}\sum_{z\in\zSet} \beta^{m,(T)}_{a,z} \KLD{P^{a,z}_{m}}{P^{a,z}_{\theta}}\rbp
\end{align*}
since $\pi_{m|m}\ra 1$ as $T\ra\infty$ and $h(\pi_{m|m})\leq\log(2)$.

Now, since $\mathcal{C}$ is a compact set for any index set $\mathcal{I}$, by the Bolzano-Weierstrass Theorem, there exists a convergent subsequence of $\lbp\bbeta^{m,(T)}\,\mv\,T\ge1\rbp$ indexed by some $\lbp T_k\rbp_{k\in{\mathbb{N}_+}}$, i.e., $\lbp\bbeta^{m,(T_k)}\,\mv\,k\in{\mathbb{N}_+}\rbp$, converging to a $\bbeta^{m,*}\in\mathcal{C}$. Furthermore, because the liminf over a sequence is upper bounded by the liminf over any of its subsequences, we have $\forall\,m\in\mSet$, $\forall\,\theta\in\mSet\setminus\{m\}$,
\begin{align*}
    &\liminf_{T\to\infty}\lbp\frac{1}{T}\log\frac{1}{\pi_{m|\theta}}\rbp\\
    &\leq\liminf_{k\to\infty}\lbp\sum_{a\in\aSet}\sum_{z\in\zSet} \beta^{m,(T_k)}_{a,z} \KLD{P^{a,z}_{m}}{P^{a,z}_{\theta}}\rbp\\
    &=\sum_{a\in\aSet}\sum_{z\in\zSet} \beta^{m,*}_{a,z} \KLD{P^{a,z}_{m}}{P^{a,z}_{\theta}},
\end{align*}
where the equality holds true since $\bbeta^{m,(T_k)}$ converges to $\bbeta^{m,*}$, and those KL-divergences are constant. 

Consequently, for any achievable $M(M-1)$ types of error exponents $\Big\{ e_{m\vert\theta} \,\Big\vert\, m\in\mSet, \theta\in\mSet\setminus\{m\}\Big\}$, we have shown that $\forall\,m\in\mSet$, there exists a $\bbeta^m\in\mathcal{C}$ ($\bbeta^{m,*}$ here) such that $\forall\, \theta\in\mSet\setminus\{m\}$,
\[
     e_{m\vert\theta}\leq\sum_{a\in\aSet}\sum_{z\in\zSet}\beta^m_{a,z} \KLD{P^{a,z}_{m}}{P^{a,z}_{\theta}},
\]
thereby completing the proof. \hfill \IEEEQED

\section{Proof of Lemma~\ref{thm:lemma1}}\label{sec:proof_of_lemma1}
Without loss of generality, let us focus on the case where hypothesis $\theta$ is the ground truth. Define the following two sets for any time slot $t\ge1$ as in the proof of \cite[Proposition 2]{NaghshvarJavidi_13}:
\[
\begin{aligned}
    &\mathbf{Q}_{\theta}(t)\eqDef\lbp\info_t\,:\,\min_{m\neq\theta}\lbp S_{t,\theta,m}-\lp TA_{\theta,m}-\epsilon_\theta\rp\rbp<0\rbp\\
    &\Tilde{\mathbf{Q}}_\theta(t)\eqDef\lbp\info_t\,:\,\min_{m\neq\theta}S_{t,\theta,m}<0\rbp
\end{aligned}
\]
where $\info_t$ given in Definition~\ref{def:decision_maker} denotes the information accumulated up to the end to time $t$. 
Furthermore, define $\Tilde{\tau}_\theta$ as follows:
\[
    \Tilde{\tau}_\theta\eqDef\inf\lbp t\,:\, \min_{m\neq\theta}S_{t',\theta, m}\geq0\,,\forall\,t'\geq t\rbp.
\]
Note that for any time slot $t\geq\Tilde{\tau}_\theta$, the maximum likelihood estimate $\hat{\theta}_{\text{MLE}}=\theta$.

We are ready to upper bound $\Pr_\theta(\tau_\theta > k)$. By the definitions of the stopping time $\tau_\theta$ and the region $\mathbf{Q}_\theta(k)$ we have: $\forall\,k\geq\kappa_{\theta,T,l}$,
\begin{align*}
    &\Pr_\theta\lp\tau_\theta>k\rp
    \le\Pr_\theta\lp \mathbf{Q}_{\theta}(k)\rp\\
    &=\Pr_{\theta}\lp\mathbf{Q}_{\theta}(k)\cap\lbp\Tilde{\tau}_\theta\le g_{k,l}\rbp\rp+\Pr_{\theta}\lp\mathbf{Q}_{\theta}(k)\cap\lbp\Tilde{\tau}_\theta>g_{k,l}\rbp\rp\\
    &\le\Pr_{\theta}\lp\mathbf{Q}_{\theta}(k)\cap\lbp\Tilde{\tau}_\theta\le g_{k,l}\rbp\rp+\Pr_{\theta}\lp\Tilde{\tau}_\theta>g_{k,l}\rp\\
    &\le\underbrace{\Pr_{\theta}\lp\mathbf{Q}_{\theta}(k)\cap\lbp\Tilde{\tau}_\theta\le g_{k,l}\rbp\rp}_{(i)}+\sum_{t'=g_{k,l}}^{\infty}\underbrace{\Pr_\theta\lp\Tilde{\mathbf{Q}}_\theta(t')\rp}_{(ii)}
\end{align*}
where $g_{k,l}\eqDef k\lp\frac{l/2}{1+l}\rp, \forall\,k\ge 1$. Recall that $l$ is a specific sequence defined in Table~\ref{table:Notations}. 

In the following, we further derive upper bounds for the term $(i)$ and the term $(ii)$ respectively.

\subsection{Upper bounding the term (i)}
For the upper bound of the term $(i)$, we slightly modify the achievability proof in \cite[Theorem 4]{Vaidhiyan2015}. Specifically, by union bound and the definition of $\mathbf{Q}_\theta(k)$, we have
\begin{align*}
    (i)
    &=\Pr_{\theta}\lp\!\!\begin{array}{l}
    \lbp\min_{m\neq\theta}\lbp S_{k,\theta,m}-\lp TA_{\theta,m}-\epsilon_\theta\rp\rbp<0\rbp\\
    \bigcap\lbp\Tilde{\tau}_\theta\le g_{k,l}\rbp
    \end{array}\!\!\rp\\
    &\le\sum_{m\neq\theta}\Pr_\theta\lp
    \lbp S_{k,\theta,m}<TA_{\theta,m}-\epsilon_\theta\rbp
    \cap\lbp\Tilde{\tau}_\theta\le g_{k,l}\rbp
    \rp\\
    &=\sum_{m\neq\theta}\Pr_\theta\lp\!\!\begin{array}{l} 
    \lbp U_k< \lp\begin{subarray}{l}TA_{\theta,m}-\epsilon_\theta\\- \sum_{s=1}^{k}\E_\theta\lb \text{LLR}^{A_s,Z_s}_{\theta,m}(\mathbf{X}_s)\,\mv\,\mathcal{F}_{s-1}\rb\end{subarray}\rp \rbp\\
    \bigcap \lbp \Tilde{\tau}_\theta\le g_{k,l} \rbp 
    \end{array}\!\!\rp,
\end{align*}
where 
\begin{align}
\notag
&\text{LLR}^{A_s,Z_s}_{\theta,m}(\mathbf{X}_s)\eqDef\log\frac{P^{A_s,Z_s}_{\theta}(\mathbf{X}_{s})}{P^{A_s,Z_s}_{m}(\mathbf{X}_{s})}\quad\forall\,s\ge1,\\
&U_k\eqDef S_{k,\theta,m}-\sum_{s=1}^{k}\E_\theta\lb\text{LLR}^{A_s,Z_s}_{\theta,m}(\mathbf{X}_s)\,\mv\,\mathcal{F}_{s-1}\rb\label{eq:U_k_def}.
\end{align}

To further bound the term $(i)$, we proceed by deriving a lower bound for $\sum_{s=1}^{k}\E_\theta\lb \text{LLR}_{\theta,m}(s)\,\mv\,\mathcal{F}_{s-1}\rb$,
\begin{align}          
\notag    &\sum_{s=1}^{k}\E_\theta\lb\text{LLR}^{A_s,Z_s}_{\theta,m}(\mathbf{X}_s)\,\mv\,\mathcal{F}_{s-1}\rb\\
\notag    &=\sum_{s=1}^{\Tilde{\tau}_\theta}\E_\theta\lb\text{LLR}^{A_s,Z_s}_{\theta,m}(\mathbf{X}_s)\,\mv\,\mathcal{F}_{s-1}\rb\\
\notag    &\quad+\sum_{s=\Tilde{\tau}_\theta+1}^{k}\E_\theta\lb \text{LLR}^{A_s,Z_s}_{\theta,m}(\mathbf{X}_s)\,\mv\,\mathcal{F}_{s-1}\rb\\
    &\ge\sum_{s=\Tilde{\tau}_\theta+1}^{k}\E_\theta\lb \text{LLR}^{A_s,Z_s}_{\theta,m}(\mathbf{X}_s)\,\mv\,\mathcal{F}_{s-1}\rb\label{eq:E_S_K}\\
    &=\sum_{s=\Tilde{\tau}_\theta+1}^{k}\sum_{a,z}P_{A_s, Z_s|\mathcal{F}_{s-1}}\lp a, z\,\mv\,\mathcal{F}_{s-1}\rp \KLD{P^{a,z}_{\theta}}{P^{a,z}_{m}}\label{eq:E_S_K_1}\\
    &=\sum_{s=\Tilde{\tau}_\theta+1}^{k}\sum_{a,z}P_{A_s|Z_s, \info_{s-1}}\lp a\,\mv\,z, f_{s-1}\rp \alpha_z \KLD{P^{a,z}_{\theta}}{P^{a,z}_{m}}\label{eq:E_S_K_2}\\
    &=\sum_{s=\Tilde{\tau}_\theta+1}^{k}\sum_{a,z}\lp \lp1-\epsilon(T)\rp \beta^{\theta}_{a,z} + \epsilon(T)\frac{1}{h}\rp \KLD{P^{a,z}_{\theta}}{P^{a,z}_{m}}\label{eq:E_S_K_3}.
\end{align}
Here \eqref{eq:E_S_K} and \eqref{eq:E_S_K_1} hold because $\forall\,s\ge1$, we have
\begin{align*}
    &\E_\theta\lb \text{LLR}^{A_s,Z_s}_{\theta,m}(\mathbf{X}_s)\,\mv\,\mathcal{F}_{s-1}\rb\\
    &=\sum_{a,z}\lp\!\!\begin{array}{l}
    P_{A_s,Z_s|\mathcal{F}_{s-1}}\lp a, z\,\mv\,\mathcal{F}_{s-1}\rp\\
    \cdot\E_\theta\lb \log\frac{P^{a,z}_{\theta}(\mathbf{X}_{s})}{P^{a,z}_{m}(\mathbf{X}_{s})} \,\mv\, A_s=a,Z_s=z, \mathcal{F}_{s-1}\rb
    \end{array}\!\!\rp\\
    &=\sum_{a,z}P_{A_s,Z_s|\mathcal{F}_{s-1}}\lp a, z\,\mv\,\mathcal{F}_{s-1}\rp\KLD{P^{a,z}_{\theta}}{P^{a,z}_{m}}
    \ge0.
\end{align*}
\eqref{eq:E_S_K_2} holds since the current available set $Z_s$ and the past information $\info_{s-1}$ are independent and we switch from conditioning on $\mathcal{F}_{s-1}$ to merely conditioning on $\info_{s-1}$ by Definition~\ref{def:decision_maker}. \eqref{eq:E_S_K_3} is due to  our proposed test in the case of $\hat{\theta}_{\text{MLE}}=\theta$ and by the definition of $\Tilde{\tau}_\theta$. 

By the definitions of $A_{\theta,m}$ and $\tilde{A}_{\theta,m}$ in Table~\ref{table:Notations}, \eqref{eq:E_S_K_3} can be expressed in an alternative form as follows:
\[
\eqref{eq:E_S_K_3} = A_{\theta,m}\lp 1+l\rp \lp k-\Tilde{\tau}_\theta\rp,
\]
where $l$ is a specific sequence also defined in Table~\ref{table:Notations}.

Consequently, the upper bound of the term $(i)$ becomes
\begin{align}
\notag    
&(i)\\
\notag
&\le\sum_{m\neq\theta}\Pr_\theta\lp\!\!\begin{array}{l} 
\lbp U_k<TA_{\theta,m}-\epsilon_\theta - A_{\theta,m}\lp 1+l\rp \lp k-\Tilde{\tau}_\theta\rp \rbp\\\bigcap \lbp \Tilde{\tau}_\theta\le g_{k,l}\rbp\end{array}\!\!\rp\\
&\le\sum_{m\neq\theta}\Pr_\theta\lp\!\!\begin{array}{l}
\lbp U_k<TA_{\theta,m}-\epsilon_\theta-k\lp 1+\frac{l}{2} \rp A_{\theta,m}\rbp\\\bigcap\lbp\Tilde{\tau}_\theta\le g_{k,l}\rbp\end{array}\!\!\rp\label{eq:k-tau}\\
&\le \sum_{m\neq\theta}\Pr_\theta\Bigg(U_k<A_{\theta,m}\underbrace{\lp T-\frac{\epsilon_\theta}{A^{\text{max}}_{\theta}}- k\lp 1+\frac{l}{2}\rp\rp}_{(*)}\Bigg), 
\label{eq:k-tau-1}
\end{align}
where \eqref{eq:k-tau} is because $\Tilde{\tau}_\theta\le g_{k,l}=k\lp\frac{l/2}{1+l}\rp$ and $A^{\text{max}}_\theta\eqDef\max_{m\neq\theta}A_{\theta,m}$ (see Table~\ref{table:Notations}).

To further  bound \eqref{eq:k-tau-1}, we leverage Azuma's inequality (see the fact below) to bound each term in the summation.
\begin{fact}[Azuma's inequality]\label{fact:azuma}
    Let $\{V_i\}_{i\ge0}\eqDef\lbp V_0, V_1, ...\rbp$ be a martingale. Furthermore, for any $i\ge1$, if $\underline{V}_i\leq V_{i}-V_{i-1}\leq \overline{V}_i$ and there exists a finite sequence $\{C_i\}_{i\ge1}\eqDef\lbp C_1, C_2,...\rbp$ such that $\overline{V}_i-\underline{V}_i\leq C_i$ for any $i\ge1$, then we have for any $v>0$, 
\[\textstyle
    P\lp V_i-V_0\leq -v\rp\leq\exp\lp\frac{-2v^2}{\sum_{s=1}^{i}C^2_s}\rp.
\]
\end{fact}

In the following, we shall first show that $\{U_i\}_{i\ge0}$ as defined in \eqref{eq:U_k_def} is a martingale. Then we identify the parameters $\{C_i\}_{i\ge 1}$ and $v$ in Azuma's inequality (Fact~\ref{fact:azuma}) to establish an upper bound of 
\[
\Pr_\theta\lp    U_k<A_{\theta,m}\lp T-\frac{\epsilon_\theta}{A^{\text{max}}_{\theta}}- k\lp 1+\frac{l}{2}\rp\rp\rp.
\]
\begin{enumerate}
\item 
To show that $\{U_i\}_{i\ge0}$ is a martingale, note that by plugging the expression of $S_{k,\theta,m}$ shown in Lemma~\ref{le:S_is_martingale} into the definition of $U_i$ in \eqref{eq:U_k_def}, we have
\begin{align}
\notag&U_i = \\
&\begin{cases}
\sum\limits_{s=1}^{i}\lp\!\!\begin{array}{l}\text{LLR}^{A_s,Z_s}_{\theta,m}(\mathbf{X}_s)\\-\E_\theta\lb \text{LLR}^{A_s,Z_s}_{\theta,m}(\mathbf{X}_s)\,\mv\,\mathcal{F}_{s-1}\rb\end{array}\!\!\rp, 
&i\ge1.\\[1ex]
0, 
&i = 0.
\end{cases}.
\label{eq:U_i_def}
\end{align}
The martingale property of the stochastic process $\{U_i\}_{i\ge0}$ can then be shown as follows:
\begin{align*}
    &\E_\theta\lb U_i-U_{i-1}\,\mv\,\mathcal{F}_{i-1}\rb\\
    &=\E_\theta\lb\!\!\begin{array}{l} 
    \text{LLR}^{A_i,Z_i}_{\theta,m}(\mathbf{X}_i)\\
    -\E_\theta\lb \text{LLR}^{A_i,Z_i}_{\theta,m}(\mathbf{X}_i)\,\mv\,\mathcal{F}_{i-1}\rb\end{array}\!\!\,\mv\,\mathcal{F}_{i-1}\rb\\
    &=\E_\theta\lb \text{LLR}^{A_i,Z_i}_{\theta,m}(\mathbf{X}_i)\,\mv\,\mathcal{F}_{i-1} \rb \\
    &\quad- \E_\theta\lb \E_\theta\lb \text{LLR}^{A_i,Z_i}_{\theta,m}(\mathbf{X}_i)\,\mv\,\mathcal{F}_{i-1}\rb\,\mv\,\mathcal{F}_{i-1}\rb\\
    &=\E_\theta\lb \text{LLR}^{A_i,Z_i}_{\theta,m}(\mathbf{X}_i)\,\mv\,\mathcal{F}_{i-1} \rb \\
    &\quad-\E_\theta\lb \text{LLR}^{A_i,Z_i}_{\theta,m}(\mathbf{X}_i)\,\mv\,\mathcal{F}_{i-1} \rb\\
    &=0.
\end{align*}

\item 
To identify the parameters $\{C_i\}_{i\ge 1}$, first note that by Assumption~\ref{a:oneStepLLR}, it is straightforward that 
\[
	|\text{LLR}^{a,z}_{\theta,m}(\mathbf{x})|=\lba \log\frac{P^{a,z}_\theta(\mathbf{x})}{P^{a,z}_m(\mathbf{x})} \rba\leq L
\]
for all $a\in\aSet$, $z\in\zSet$, $\theta,m\in\mSet$, $\theta\neq m$, $\mathbf{x}\in\mathcal{X}^{a,z}_\theta$, where $\mathcal{X}^{a,z}_\theta$ stands for the support of $P^{a,z}_\theta$. 
Hence, we have $\forall\,i\ge1$,
\[
    -2   L \le U_{i}-U_{i-1}\le 2   L  
\]
where 
\[\begin{aligned}
&U_{i}-U_{i-1}\\&=\text{LLR}^{A_i,Z_i}_{\theta,m}(\mathbf{X}_i)-\E_\theta\lb \text{LLR}^{A_i,Z_i}_{\theta,m}(\mathbf{X}_i)\,\mv\,\mathcal{F}_{i-1}\rb.
\end{aligned}
\]
As a result, by assigning $\ol{V}_i = 2 L$ and $\ul{V}_i = -2 L$, we have $\ol{V}_i-\ul{V}_i\le 4  L$ and hence choose $C_i =4 L$, $\forall\,i\geq1$.

\item
We aim to identify the parameter $v$ as 
\[
    v=A_{\theta,m}\lp k\lp 1+\frac{l}{2}\rp + \frac{\epsilon_\theta}{A^{\text{max}}_{\theta}} -T\rp.
\]
To do so, we just need to show that the term $(*)$ defined in \eqref{eq:k-tau-1} times $A_{\theta,m}$ is smaller than zero.

First note that by Assumption~\ref{a:discrimination_1}, we have $\Tilde{A}_{\theta,m}>0$, namely, 
\[
A_{\theta,m}>0\quad \forall\,\theta,m\in\mSet, \theta\neq m.
\]

It remains to show that the term $(*)<0$. When $k\geq\kappa_{\theta,T,l}=\lceil T-\frac{\epsilon_\theta}{A^{\text{max}}_\theta} \rceil$, we have $\forall\,T\geq q\lp l,n,I(T)\rp$,
\begin{align}
\notag    (*)&\le T-\frac{\epsilon_\theta}{A^{\text{max}}_\theta}-\lp 1+\frac{l}{2}\rp\kappa_{\theta,T,l}\\
\notag    &\le T-\frac{\epsilon_\theta}{A^{\text{max}}_\theta}-\lp1+\frac{l}{2}\rp\lp T-\frac{\epsilon_\theta}{A^{\text{max}}_\theta}\rp\\
    &=-\frac{l}{2}\lp T-\frac{\epsilon_\theta}{A^{\text{max}}_\theta}\rp< 0\label{eq:T-ep/A}
\end{align}
where \eqref{eq:T-ep/A} holds because $l\ge0$ and by the following derivation:
\begin{align}
\notag    T-\frac{\epsilon_\theta}{A^{\text{max}}_{\theta}}&=T-(1-\frac{K(\Delta)}{b(l)})\\
    &\geq T-1-\frac{1}{b(l)} \label{eq:extra_1}\\
\notag    &\geq q\lp l,n,I(T)\rp-1-\frac{1}{b(l)}\\
\notag    &=\frac{\log\left(MB(l)(1+b(l))\right)}{b(l)}\\
\notag    &>0\quad\forall\,T\geq q\lp l,n,I(T)\rp. 
\end{align}
Note that in the derivation above, we plug in the definitions of $\epsilon_\theta$, $q\lp l,n,I(T)\rp$, $\Delta$, $b(l)$, $B(l)$, and $K(\cdot)$ as shown in Table~\ref{table:Notations}. 
More specifically, \eqref{eq:extra_1} holds because as $T\geq q(l,n,I(T))$, we have $\Delta\ge -e^{-1}$ and hence $-1\le K(\Delta)\le0$ by the range of the function $K(\cdot)$.
\end{enumerate}

In summary, we apply Fact~\ref{fact:azuma} with $C_i = 4 L\ \forall\,i\ge1$ and $v=A_{\theta,m}\lp k\lp 1+\frac{l}{2}\rp + \frac{\epsilon_\theta}{A^{\text{max}}_{\theta}} -T\rp$ and get
\[
\begin{aligned}
&\Pr_\theta\lp U_k<A_{\theta,m}\lp T-\frac{\epsilon_\theta}{A^{\text{max}}_{\theta}}- k\lp 1+\frac{l}{2}\rp\rp\rp \\
&\le 
\exp\lp\frac{-2A^2_{\theta,m}\lp k\lp 1+\frac{l}{2}\rp + \frac{\epsilon_\theta}{A^{\text{max}}_{\theta}} - T \rp^2}{k\lp 4 L\rp^2}\rp.
\end{aligned}
\]
Consequently, 
\begin{align}
\notag (i)&\le\sum_{m\neq\theta}\exp\lp\frac{-2A^2_{\theta,m}\lp k\lp 1+\frac{l}{2}\rp + \frac{\epsilon_\theta}{A^{\text{max}}_{\theta}} - T \rp^2}{k\lp 4 L\rp^2}\rp\\
\notag &\le\sum_{m\neq\theta}\exp\lp \frac{-A^2_{\theta,m}k\lp\lp1+\frac{l}{2}\rp +\frac{1}{k} \lp \frac{\epsilon_\theta}{A^{\text{max}}_{\theta}}-T \rp \rp^2}{8 L^2} \rp\\
&\leq\sum_{m\neq\theta}e^{-\frac{A^2_{\theta,m}k(\frac{l}{2})^2}{8 L^2}} \label{eq:extra_3}\\
&\leq Me^{-k\frac{l^2}{2}(\frac{\min_{m\neq\theta}A_{\theta,m}}{4 L})^2} \label{eq:uppbd_part(i)},
\end{align}
where \eqref{eq:extra_3} holds because $\forall\,k\geq\kappa_{\theta,T,l}\eqDef \left\lceil T-\frac{\epsilon_\theta}{A^{\text{max}}_\theta} \right\rceil$, $1+ \frac{1}{k}\lp\frac{\epsilon_\theta}{A^{\text{max}}_{\theta}}-T\rp\ge 0$.

\subsection{Upper bounding the term $(ii)$}
To derive the upper bound for the term $(ii)$, let us begin with
\begin{align}
    (ii)&=\Pr_\theta\lp\Tilde{\mathbf{Q}}_\theta(t')\rp\le\sum_{m\neq\theta}\Pr_\theta\lp S_{t',\theta,m}<0\rp\label{eq:extra_4}\\
    &=\sum_{m\neq\theta}\Pr_\theta\lp U_{t'}<-\sum_{s=1}^{t'} \E_\theta\lb \text{LLR}^{A_s,Z_s}_{\theta,m}(\mathbf{X}_s)\,\mv\,\mathcal{F}_{s-1}\rb \rp\label{eq:extra_4_1}.
\end{align}
Here \eqref{eq:extra_4} holds true by union bound and the definition of $\Tilde{\mathbf{Q}}_\theta(t')$. In \eqref{eq:extra_4_1}, recall the form of $U_{t'}$ in \eqref{eq:U_i_def}. 

To proceed, we have $\forall\,s\ge1$, 
\begin{align}
\notag
&\E_\theta\lb \text{LLR}^{A_s,Z_s}_{\theta,m}(\mathbf{X}_s)\,\mv\,\mathcal{F}_{s-1}\rb\\
\notag
&=\sum_{a,z}P_{A_s|Z_s, \info_{s-1}}\lp a\,\mv\,z, f_{s-1}\rp \alpha_z \KLD{P^{a,z}_{\theta}}{P^{a,z}_{m}}\\
&\ge\sum_{a,z}\epsilon(T)\frac{1}{h}\KLD{P^{a,z}_{\theta}}{P^{a,z}_{m}}\label{eq:extra_5}\\
\notag
&\ge \epsilon(T)\min_{\theta\in\mSet}\min_{m\neq\theta}\frac{1}{h}\sum_{a,z}\KLD{P^{a,z}_{\theta}}{P^{a,z}_{m}}\\
\notag
&= \epsilon(T)\min_{\theta\in\mSet}\min_{m\neq\theta}\Tilde{A}_{\theta,m} = I(T),
\end{align}
where the last line follows from the definition of $\Tilde{A}_{\theta,m}$ and $I(T)$ in Table~\ref{table:Notations}, and 
\eqref{eq:extra_5} follows directly from our proposed test: $\forall\,a\in\aSet\setminus\{\emptyset\}$, $\forall\,z\in\zSet$, $\forall\,f_{s-1}$, 
\[
P_{A_s|Z_s, \info_{s-1}}\lp a\,\mv\,z, f_{s-1}\rp\ge\epsilon(T)\frac{1}{h\alpha_z}.
\]

Hence, \eqref{eq:extra_4_1} is further upper bounded as
\[
    \eqref{eq:extra_4_1} \le \sum_{m\neq\theta}\Pr_\theta\lp U_{t'}<-t'I(T)\rp.
\]
Again, applying Azuma's inequality (Fact~\ref{fact:azuma}) with $v = t'I(T)$ and $C_i = 4L\ \forall\,i\ge1$, we have
\begin{align}
\notag    (ii)&\le M e^{-t'\frac{I^2(T)}{8 L^2}}\\
    &\le M e^{-t'\frac{l^2}{2}(\frac{I(T)/(1+l/2)}{4  L})^2}\label{eq:term_ii_2},
\end{align}
where \eqref{eq:term_ii_2} holds because 
\[\textstyle
\frac{I^2\lp T\rp}{8 L^2} 
\ge \frac{I^2\lp T\rp}{8 L^2} \lp \frac{l}{2\lp 1+\frac{l}{2}\rp} \rp^2=\frac{l^2}{2}\lp\frac{I\lp T\rp/\lp 1+l/2\rp}{4  L}\rp^2.
\] 

\subsection{Combining the upper bounds of the term $(i)$ and $(ii)$}
Finally, the upper bound of $\Pr_\theta(\tau_\theta>k)$ can be obtained by combining the upper bound of the term $(i)$ in \eqref{eq:uppbd_part(i)} and that of the term $(ii)$ in \eqref{eq:term_ii_2} as follows,
\begin{align}
\notag
&\Pr_\theta\lp \tau_\theta>k\rp\\
\notag
&\le 
    Me^{-k\frac{l^2}{2}(\frac{\min_{m\neq\theta}A_{\theta,m}}{4 L})^2} 
    + \sum_{t'=g_{k,l}}^{\infty}M e^{-t'\frac{l^2}{2}(\frac{I(T)/(1+l/2)}{4  L})^2}\\
\notag
&= 
    Me^{-k\frac{l^2}{2}(\frac{\min_{m\neq\theta}A_{\theta,m}}{4  L})^2}
    + M\frac{e^{-k\frac{l^3}{4(1+l)}(\frac{I(T)/(1+l/2)}{4  L})^2}}{1-e^{-\frac{l^2}{2}(\frac{I(T)/(1+l/2)}{4  L})^2}}\\
&\le Me^{-k\frac{l^2}{2(1+l)^2}(\frac{D}{4  L})^2}+ M\frac{e^{-k\frac{l^3}{4(1+l)^3}(\frac{I(T)}{4  L})^2}}{1-e^{-\frac{l^2}{2(1+l)^2}(\frac{I(T)}{4  L})^2}},
    \label{eq:final_upperbd}
\end{align}
where $D\eqDef\min_{\theta\in\mSet}\min_{m\neq\theta}A_{\theta,m}$. 

Subsequently, by the definitions of $l$, $I(T)$, and $\epsilon(T)$ (see Table~\ref{table:Notations}), we have $\forall\,T\geq e$,
\begin{align*}
D^2
&=lT^{\frac{1}{6}}\lp\frac{I(T)}{\epsilon(T)}\rp^2
=T^{\frac{1}{6}}l \lp\log(T)\rp^{\frac{1}{2}}I^2(T)\\
&>I^2(T)l>I^2(T)\frac{l}{2(1+l)},
\end{align*}
where $T^{1/6}\lp\log\lp T\rp\rp^{1/2}>1$, $\forall\,T\ge e$. 
Hence, it is straightforward that
\begin{equation}\label{eq:extra_6}
    k\frac{l^2}{2\lp 1+l\rp^2}\lp\frac{D}{4  L}\rp^2>k\frac{l^3}{4(1+l)^3}\lp\frac{I(T)}{4  L}\rp^2.
\end{equation}

Then, plugging \eqref{eq:extra_6} back into \eqref{eq:final_upperbd} and by the definitions of $b(l)$ and $B(l)$ (see Table~\ref{table:Notations}), we have
\begin{align}
\notag    &\Pr_\theta\lp\tau_\theta>k\rp\\
\notag    &\le Me^{-k\frac{l^3}{4(1+l)^3}\lp\frac{I(T)}{4 L}\rp^2}\lp1+\frac{1}{1-e^{-\frac{l^2}{2(1+l)^2}\lp\frac{I(T)}{4 L}\rp^2}}\rp\\
    &\le Me^{-kb(l)}\lp2+\frac{1}{\frac{l^2}{2(1+l)^2}\lp\frac{I(T)}{4  L}\rp^2}\rp\label{eq:extra_7}\\ 
\notag    &=Me^{-kb(l)}B(l),
\end{align}
where \eqref{eq:extra_7} holds since $\frac{1}{1-e^{-x}}\le 1+\frac{1}{x},\forall\,x>0$. This completes the proof of Lemma~\ref{thm:lemma1}.
\hfill \IEEEQED

\section{Conclusion}\label{sec:conclusion}
In this work, we characterize the optimal individual error exponent region in active sequential multiple-hypothesis testing problems with temporally available data sources under linear expected budget constraints (including the expected stopping time constraint). From the characterization of the region, it is found that the tradeoffs among the individual error exponents only appear when there are more than two hypotheses to be distinguished. Such tradeoffs are mainly due to the heterogeneous level of informativeness across data sources in distinguishing a declared hypothesis against other ground truth hypotheses, and they appear even in the most basic setting in \cite{Chernoff_59}. 
Our asymptotically optimal test strikes a balance between exploitation and exploration and enjoys a less strong assumption than Assumption~\ref{a:discrimination_strong} while achieving the optimal error exponents. 
The main result is general and encompasses various scenarios in the existing literature, leading to the characterization of performance gains in terms of different metrics due to adaptivity. For future work, one fruitful direction is to extend our results on individual error exponents to settings with unknown distributions of each data sources, namely, the composite hypothesis testing problem, where the state of the art for asymptotic optimality only considers the single performance metric (the maximal error probability, that is, the probability of false detection) versus a total cost \cite{PrabhuBhashyam_22}. While the general composite hypothesis setting might be intractable, it may worth the effort to focus on special cases such as the outlier hypothesis testing problem \cite{LiVeeravalli_17,ZhouWei_22}.

\appendix
\subsection{Proof of Lemma~\ref{le:S_is_martingale}}\label{app:pf_lem_S_is_martingale}
\begin{IEEEproof}
We first derive the explicit form of $\msf{P}_{m}^{(t)}$ for any $m\in\mSet$ and any $t\geq 1$, demonstrating that $S_{t,m,\theta}$ can be expressed as the sum of log-likelihood ratios. 
Subsequently, we show the martingale property by leveraging the additivity of log-likelihood ratios.

Firstly, let us derive the explicit form of $S_{t,m,\theta}$. We start with the following expression:
$\forall\,t\geq1$, $\forall\,f_t$, $\forall\,m\in\mSet$,
\begin{align}
\notag
&\msf{P}_{m}^{(t)}\lp F_t=f_t\rp\\
\notag
&= \Pr_m\lp\mathbf{X}_1=\mathbf{x}_1,Z_1=z_1,A_1=a_1\rp\\
\notag
&\quad\cdot\prod_{s=2}^{t}\Pr_m\lp\mathbf{X}_s=\mathbf{x}_s, Z_s=z_s, A_s=a_s\,\mv\,F_{s-1}=f_{s-1}\rp\\
\notag
&=\Pr_m\lp\mathbf{X}_1=\mathbf{x}_1,Z_1=z_1,A_1=a_1\rp\\
&\quad\cdot\prod_{s=2}^{t}P_Z\lp z_s\rp\Pr_m\lp\mathbf{X}_s=\mathbf{x}_s, A_s=a_s\,\mv\,\begin{subarray}{l}F_{s-1}=f_{s-1},\\Z_s=z_s\end{subarray}\rp\label{eq:express_2}\\
\notag
&=\Pr_m\lp\mathbf{X}_1=\mathbf{x}_1,Z_1=z_1,A_1=a_1\rp\\
\notag
&\quad\cdot\prod_{s=2}^{t}\lp\!\!\begin{array}{l}
P_Z\lp z_s\rp P_{A_s|Z_s, \info_{s-1}}\lp a_s\,\mv\, z_s, f_{s-1}\rp\\
\cdot\Pr_m\lp\mathbf{X}_s=\mathbf{x}_s\,\mv\,\begin{subarray}{l}
F_{s-1}=f_{s-1},Z_s=z_s,\\A_s=a_s\end{subarray}\rp
\end{array}\!\!\rp\\
\notag
&=p_{\text{init}}\prod_{s=2}^{t}P_Z\lp z_s\rp P_{A_s|Z_s, \info_{s-1}}\lp a_s\,\mv\, z_s, f_{s-1}\rp\\
&\quad\cdot\prod_{s=1}^{t}P^{a_s,z_s}_{m}\lp \mathbf{x}_{s} \rp \label{eq:express_3}
\end{align}
where $p_{\text{init}}\eqDef P_Z(z_1)P_{A_1|Z_1}(a_1\,\vert\,z_1)$, $\mathbf{x}_t\eqDef\{x_{t,j}\,\vert\, j\in a_t\cap z_t\}$, and by convention, $\msf{P}_{m}^{(0)}\lp f_0\rp\eqDef0$. Note that \eqref{eq:express_2} holds true since $Z_t$ and $F_{t-1}$ are independent $\forall\,t\geq1$, and \eqref{eq:express_3} is due to the conditional independence of samples given the current actions across time slots and the distribution of $\mathbf{X}_s$ is $P^{a_s,z_s}_{m}$ given $A_s=a_s, Z_s=z_s$ under hypothesis $m$, $\forall\,s\ge1$.

Hence, we have the explicit expression of $S_{t,m,\theta}$: $\forall\,m,\theta\in\mSet$, $m\neq\theta$, $\forall\,t\geq1$, 
\[
	S_{t,m,\theta}=\log{\frac{\msf{P}_{m}^{(t)}(\info_t)}{\msf{P}_{\theta}^{(t)}(\info_t)}}=\sum_{s=1}^{t}\log{\frac{P^{A_s,Z_s}_{m}(\mathbf{X}_{s})}{P^{A_s,Z_s}_{\theta}(\mathbf{X}_{s})}}
\]
where $S_{0,m,\theta}\eqDef 0$ by convention. 

Then, we are ready to show the martingale property. To begin with, $\forall\,t\geq0$, by the expression of $S_{t,m,\theta}$, we have  
\[
\mathbb{E}_\theta\lb e^{S_{t+1,m,\theta}}\,\mv\, \mathcal{F}_{t}\rb=e^{S_{t,m,\theta}}\mathbb{E}_\theta\lb e^{\log{\frac{P^{A_{t+1},Z_{t+1}}_{m}(\mathbf{X}_{t+1})}{P^{A_{t+1},Z_{t+1}}_{\theta}(\mathbf{X}_{t+1})}}}\, \mv\,  \mathcal{F}_{t}\rb
\]
where $\lbp e^{S_{t,m,\theta}}\,\mv\,t\geq0\rbp$ is adapted to the natural filtration $\lbp \mathcal{F}_{t}\rbp_{t\geq0}$. 
The goal is turned to show for any $t\ge0$,
\[
    \mathbb{E}_\theta\lb e^{\log{\frac{P^{A_{t+1},Z_{t+1}}_{m}(\mathbf{X}_{t+1})}{P^{A_{t+1},Z_{t+1}}_{\theta}(\mathbf{X}_{t+1})}}}\, \mv\,  \mathcal{F}_{t}\rb=1.
\]

Since the distribution of action taking at time slot $t+1$ is a function of $\info_t$ and $Z_{t+1}$ (see Definition~\ref{def:decision_maker}), 
we simply switch from conditioning on $\mathcal{F}_t$ to merely conditioning on $\info_t$ without loss of generality in the following steps. So, we have, $\forall\,t\ge0$,
\begin{align*}
&\mathbb{E}_\theta\lb e^{\log{\frac{P^{A_{t+1},Z_{t+1}}_{m}(\mathbf{X}_{t+1})}{P^{A_{t+1},Z_{t+1}}_{\theta}(\mathbf{X}_{t+1})}}}\, \mv\,  \mathcal{F}_{t}\rb\\
&=\sum_{a\in\aSet}\sum_{z\in\zSet}
\lp\!\!\begin{array}{l}
P_{A_{t+1},Z_{t+1}|\info_{t}}\lp a, z\,\mv\,f_{t} \rp\\
\cdot\mathbb{E}_\theta\lb e^{\log{\frac{P^{a,z}_{m}(\mathbf{X}_{t+1})}{P^{a,z}_{\theta}(\mathbf{X}_{t+1})}}}\,\mv\,
\begin{subarray}{l}
A_{t+1}=a,Z_{t+1}=z,\\\info_t=f_{t}
\end{subarray}\rb
\end{array}\!\!\rp\\
&=\sum_{a\in\aSet}\sum_{z\in\zSet}
\lp\!\!\begin{array}{l}
P_{A_{t+1},Z_{t+1}|\info_{t}}\lp a, z\,\mv\,f_{t} \rp\\
\cdot\mathbb{E}_\theta\lb \frac{P^{a,z}_{m}(\mathbf{X}_{t+1})}{P^{a,z}_{\theta}(\mathbf{X}_{t+1})}\,\mv\,
\begin{subarray}{l}
A_{t+1}=a,Z_{t+1}=z,\\\info_t=f_{t}
\end{subarray}\rb
\end{array}\!\!\rp.
\end{align*}

Subsequently, since the random sample set $\mathbf{X}_{t+1}$ follows the marginal distribution $P^{a,z}_\theta$ under a given $A_{t+1}=a$ and $Z_{z+1}=z$, and the samples are conditionally independent given the current action across time slots, we have
\[
\mathbb{E}_\theta\lb \frac{P^{a,z}_{m}(\mathbf{X}_{t+1})}{P^{a,z}_{\theta}(\mathbf{X}_{t+1})}\,\mv\,A_{t+1}=a,Z_{t+1}=z,\info_t=f_{t}\rb=1
\]
by Assumption~\ref{a:oneStepLLR}.

To complete the proof, we can conclude, $\forall\,t\ge0$,
\begin{align*}
&\mathbb{E}_\theta\lb e^{\log{\frac{P^{A_{t+1},Z_{t+1}}_{m}(\mathbf{X}_{t+1})}{P^{A_{t+1},Z_{t+1}}_{\theta}(\mathbf{X}_{t+1})}}}\, \mv\,  \mathcal{F}_{t}\rb\\
&=\sum_{a\in\aSet}\sum_{z\in\zSet}P_{A_{t+1},Z_{t+1}|\info_{t}}\lp a, z\,\mv\,f_{t}\rp\\
&=\sum_{z\in\zSet}\alpha_z\sum_{a\in\aSet}P_{A_{t+1}|Z_{t+1},\info_{t}}\lp a\,\mv\,z, f_{t}\rp
=1
\end{align*}
where the last equality holds by Definition~\ref{def:decision_maker} and the stochastic model of temporally available data sources in Section~\ref{subsec: stat_model}.
\end{IEEEproof}

\subsection{Proof of Proposition~\ref{thm:polytope}}\label{app:polytope}
To prove Proposition~\ref{thm:polytope}, due to the direct product form of the region $\zeta$, it is sufficient to show that, $\forall\,m\in\mSet$,
\[\textstyle
    \bigcup\nolimits_{\bbeta^m\in\mathcal{C}}\zeta_m(\bbeta^m)=\text{Conv}\lbp\text{extr}\lp\bigcup\nolimits_{\bbeta^m\in V_c}\zeta_m(\bbeta^m)\rp\rbp
\]
Note that for notational simplicity, we let $\underline{x}\eqDef\lbp x_i\,\mv\,i\in\mSet\setminus\{m\} \rbp$ and in the following proof, we just consider a fixed index $m\in\mSet$ without loss of generality. Additionally, the definition of an extreme point is that, for any set $A$, a point $a\in A$ is an extreme point of the set $A$ if for any $b,c\in A$, $b\neq c$ and for any $\lambda\in(0,1)$, the point $a$ satisfies $a\neq\lambda b +(1-\lambda)c$.

\subsubsection*{Direction ``LHS $\subseteq$ RHS"}
For any $\underline{x}\in\bigcup_{\bbeta^m\in\mathcal{C}}\zeta_m(\bbeta^m)$, there exists a $\Tilde{\bbeta}^{m}\in\mathcal{C}$ and some $\eta_1,...,\eta_{M-1}\in\lb0,1\rb$ such that
\[
x_i = \eta_i e^*_{m\vert i}(\Tilde{\bbeta}^{m})\quad\forall\,i\in\mSet\setminus\{m\}
\]
where $e^*_{m\vert i}(\Tilde{\bbeta}^{m})$ is defined in Theorem~\ref{thm:softErrorExponent}.

Moreover, since the constraint set $\mathcal{C}$ is clearly a convex polytope as $|\mcal{I}|<\infty$, we assume that $V_c=\{\tilde{\underline{v}}_1,...,\tilde{\underline{v}}_p\}$ where $p<\infty$. Therefore, there exists a sequence of $\lambda_1,..,\lambda_p\in\lb 0,1\rb$ and $\sum_{j=1}^{p}\lambda_j=1$ such that $\Tilde{\bbeta}^m=\sum_{j=1}^{p}\lambda_j\tilde{\underline{v}}_j$. 

As a result, by the definition of $e^*_{m\vert i}(\Tilde{\bbeta}^{m})$, we have, $\forall\,i\in\mSet\setminus\{m\}$,
\[\textstyle
x_i = \eta_i e^*_{m\vert i}\lp\sum_{j=1}^{p}\lambda_j\tilde{\underline{v}}_j\rp=\sum_{j=1}^{p}\lambda_j \eta_i e^*_{m\vert i}\lp\tilde{\underline{v}}_j\rp
\]
Then, due to $\lbp \eta_i e^*_{m\vert i}\lp \tilde{v}_j\rp\,\mv\,i\in\mSet\setminus\{m\} \rbp \in\zeta_m\lp\tilde{\underline{v}}_j\rp$ for any $j\in\{1,...,p\}$, it is straightforward that
\[\textstyle
\underline{x}\in\text{Conv}\lbp\bigcup\nolimits_{\bbeta^m\in V_c}\zeta_m(\bbeta^m)\rbp
\]
Note that the index $m$ in $\bbeta^m$ here is just for the notational consistency with respect to our main theorem.

Furthermore, because the region $\text{Conv}\lbp\bigcup_{\bbeta^m\in V_c}\zeta_m(\bbeta^m)\rbp$ is a compact convex set, by Krein--Milman Theorem (a well-known result in Topology), we have 
\[\begin{aligned}
&\textstyle
\text{Conv}\lbp\bigcup\nolimits_{\bbeta^m\in V_c}\zeta_m(\bbeta^m)\rbp 
\\
&\textstyle
= \text{Conv}\lbp \text{extr}\lp \text{Conv}\lbp\bigcup\nolimits_{\bbeta^m\in V_c}\zeta_m(\bbeta^m)\rbp\rp\rbp.
\end{aligned}
\]

Consequently, by the fact that the convex hull operation on the set $\bigcup_{\bbeta^m\in V_c}\zeta_m(\bbeta^m)$ does not add or eliminate any extreme points of $\bigcup_{\bbeta^m\in V_c}\zeta_m(\bbeta^m)$, namely,
\[\textstyle
\text{extr}\lp \text{Conv}\lbp\bigcup_{\bbeta^m\in V_c}\zeta_m(\bbeta^m)\rbp\rp=\text{extr}\lp\bigcup_{\bbeta^m\in V_c}\zeta_m(\bbeta^m)\rp,
\]
we can conclude that
\[\begin{aligned}
&\textstyle
\text{Conv}\lbp \text{extr}\lp \text{Conv}\lbp\bigcup_{\bbeta^m\in V_c}\zeta_m(\bbeta^m)\rbp\rp\rbp\\
&\textstyle
=\text{Conv}\lbp \text{extr}\lp\bigcup_{\bbeta^m\in V_c}\zeta_m(\bbeta^m)\rp\rbp,
\end{aligned}
\]
thereby 
\[\begin{aligned}
&\textstyle
\bigcup_{\bbeta^m\in\mathcal{C}}\zeta_m(\bbeta^m)\subseteq\text{Conv}\lbp\bigcup_{\bbeta^m\in V_c}\zeta_m(\bbeta^m)\rbp\\
&\textstyle
=\text{Conv}\lbp\text{extr}\lp\bigcup_{\bbeta^m\in V_c}\zeta_m(\bbeta^m)\rp\rbp.
\end{aligned}
\]

\subsubsection*{Direction ``LHS $\supseteq$ RHS"}
Since $\mathcal{C}$ is a convex polytope, $V_c$ is a finite set when $|\mcal{I}|<\infty$. Hence, we assume there are $k<\infty$ extreme points of the region $\bigcup_{\bbeta^m\in V_c}\zeta_m(\bbeta^m)$, i.e., $\text{extr}\lp\bigcup_{\bbeta^m\in V_c}\zeta_m(\bbeta^m)\rp=\{\underline{v}_1,\underline{v}_2,...,\underline{v}_k\}$. 
Then, we have $\forall\,\underline{x}\in\text{Conv}\lbp\text{extr}\lp\bigcup_{\bbeta^m\in V_c}\zeta_m(\bbeta^m)\rp\rbp$, 
$\exists\lambda_1,\lambda_2,...,\lambda_k\in[0,1]$, $\sum_{i=1}^{k}\lambda_i=1$, such that 
$\underline{x}=\sum_{i=1}^{k}\lambda_i\underline{v}_i$. 
Furthermore, we know $\forall\,i\in\{1,...,k\}$, 
\[\textstyle
	\underline{v}_i\in\bigcup_{\bbeta^m\in V_c}\zeta_m(\bbeta^m)\subseteq\bigcup_{\bbeta^m\in \mathcal{C}}\zeta_m(\bbeta^m), 
\]
and $\bigcup_{\bbeta^m\in \mathcal{C}}\zeta_m(\bbeta^m)$ is also a convex set. 
Therefore, we have
\[\textstyle
	\underline{x}\in\bigcup_{\bbeta^m\in\mathcal{C}}\zeta_m(\bbeta^m).
\]
As a result,
\[\textstyle
\bigcup_{\bbeta^m\in\mathcal{C}}\zeta_m(\bbeta^m)\supseteq\text{Conv}\lbp\text{extr}\lp\bigcup_{\bbeta^m\in V_c}\zeta_m(\bbeta^m)\rp\rbp.
\]

\hfill 
\IEEEQED

\bibliographystyle{IEEEtran}

\begin{thebibliography}{10}
\providecommand{\url}[1]{#1}
\csname url@samestyle\endcsname
\providecommand{\newblock}{\relax}
\providecommand{\bibinfo}[2]{#2}
\providecommand{\BIBentrySTDinterwordspacing}{\spaceskip=0pt\relax}
\providecommand{\BIBentryALTinterwordstretchfactor}{4}
\providecommand{\BIBentryALTinterwordspacing}{\spaceskip=\fontdimen2\font plus
\BIBentryALTinterwordstretchfactor\fontdimen3\font minus
  \fontdimen4\font\relax}
\providecommand{\BIBforeignlanguage}[2]{{%
\expandafter\ifx\csname l@#1\endcsname\relax
\typeout{** WARNING: IEEEtran.bst: No hyphenation pattern has been}%
\typeout{** loaded for the language `#1'. Using the pattern for}%
\typeout{** the default language instead.}%
\else
\language=\csname l@#1\endcsname
\fi
#2}}
\providecommand{\BIBdecl}{\relax}
\BIBdecl

\bibitem{Chernoff_59}
H.~Chernoff, ``Sequential design of experiments,'' \emph{The Annals of
  Mathematical Statistics}, vol.~30, no.~3, pp. 755--770, 1959.

\bibitem{NaghshvarJavidi_13}
M.~Naghshvar and T.~Javidi, ``Active sequential hypothesis testing,'' \emph{The
  Annals of Statistics}, vol.~41, no.~6, pp. 2703--2738, 2013.

\bibitem{NitinawaratAtia_13}
S.~Nitinawarat, G.~K. Atia, and V.~V. Venugopal, ``Controlled sensing for
  multihypothesis testing,'' \emph{IEEE Transactions on Automatic Control},
  vol.~58, no.~10, pp. 2451--2464, October 2013.

\bibitem{BaiGupta_17}
C.-Z. Bai and V.~Gupta, ``An on-line sensor selection algorithm for {SPRT} with
  multiple sensors,'' \emph{IEEE Transactions on Automatic Control}, vol.~62,
  no.~7, pp. 3532--3539, July 2017.

\bibitem{LiLi_19}
S.~Li, X.~Li, X.~Wang, and J.~Liu, ``Sequential hypothesis test with online
  usage-constrained sensor selection,'' \emph{IEEE Transactions on Information
  Theory}, vol.~65, no.~7, pp. 4392--4410, July 2019.

\bibitem{BaiKatewa_15}
C.-Z. Bai, V.~Katewa, V.~Gupta, and Y.~Huang, ``A stochastic sensor selection
  scheme for sequential hypothesis testing with multiple sensors,'' \emph{IEEE
  Transactions on Signal Processing}, vol.~63, no.~14, pp. 3687--3699, July
  2015.

\bibitem{Vaidhiyan2015}
N.~K. Vaidhiyan and R.~Sundaresan, ``Active search with a cost for switching
  actions,'' \emph{2015 Information Theory and Applications Workshop (ITA)},
  pp. 17--24, 2015.

\bibitem{LanWang_21}
S.-W. Lan and I.-H. Wang, ``Heterogeneous sequential hypothesis testing with
  active source selection under budget constraints,'' \emph{IEEE International
  Symposium on Information Theory (ISIT)}, 2021.

\bibitem{DragalinTartakovsky_99}
V.~P. Dragalin, A.~G. Tartakovsky, and V.~V. Veeravalli, ``Multihypothesis
  sequential probability ratio tests -- {P}art {I}: Asymptotic optimality,''
  \emph{IEEE Transactions on Information Theory}, vol.~45, no.~7, pp.
  2448--2461, November 1999.

\bibitem{Hoeffding_65}
W.~Hoeffding, ``Asymptotically optimal tests for multinomial distributions,''
  \emph{Annals of Mathematical Statistics}, vol.~36, pp. 369--401, 1965.

\bibitem{Blahut_74}
R.~Blahut, ``Hypothesis testing and information theory,'' \emph{IEEE
  Transactions on Information Theory}, vol.~20, no.~4, pp. 405--417, 1974.

\bibitem{WaldWolfowitz_48}
A.~Wald and J.~Wolfowitz, ``Optimum character of the sequential probability
  ratio test,'' \emph{The Annals of Mathematical Statistics}, vol.~19, no.~3,
  pp. 326--339, September 1948.

\bibitem{polyanskiy2025information}
Y.~Polyanskiy and Y.~Wu, \emph{Information Theory: From Coding to
  Learning}.\hskip 1em plus 0.5em minus 0.4em\relax Cambridge University Press,
  2025.

\bibitem{Tuncel_05}
E.~Tuncel, ``On error exponents in hypothesis testing,'' \emph{IEEE
  Transactions on Information Theory}, vol.~51, no.~8, pp. 2945--2950, 2005.

\bibitem{NaghshvarJavidi_13_2}
M.~Naghshvar and T.~Javidi, ``Sequentiality and adaptivity gains in active
  hypothesis testing,'' \emph{IEEE Journal of Selected Topics in Signal
  Processing}, vol.~7, no.~5, pp. 768--782, 2013.

\bibitem{HsuWang_ISIT24}
C.-Y. Hsu and I.-H. Wang, ``On error exponents in hypothesis testing,''
  \emph{IEEE International Symposium on Information Theory (ISIT)}, 2024.

\bibitem{Roig2020}
B.~Roig-Solvas and M.~Sznaier, ``Euclidean distance bounds for linear matrix
  inequalities analytic centers using a novel bound on the lambert function,''
  \emph{SIAM Journal on Control and Optimization}, vol.~60, no.~2, pp.
  720--731, April 2022.

\bibitem{PrabhuBhashyam_22}
G.~R. Prabhu, S.~Bhashyam, A.~Gopalan, and R.~Sundaresan, ``Sequential
  multi-hypothesis testing in multi-armed bandit problems: An approach for
  asymptotic optimality,'' \emph{IEEE Transactions on Information Theory},
  vol.~68, no.~7, pp. 4790--4817, July 2022.

\bibitem{LiVeeravalli_17}
Y.~Li, S.~Nitinawarat, and V.~V. Veeravalli, ``Universal sequential outlier
  hypothesis testing,'' \emph{Sequential Analysis}, vol.~36, no.~3, pp.
  309--344, 2017.

\bibitem{ZhouWei_22}
L.~Zhou, Y.~Wei, and I.~Alfred O.~Hero, ``Second-order asymptotically optimal
  outlier hypothesis testing,'' \emph{IEEE Transactions on Information Theory},
  vol.~68, no.~6, pp. 3583--3607, June 2022.

\end{thebibliography}

\begin{IEEEbiographynophoto}{Chia-Yu Hsu}
(Member, IEEE) received B.S. degree in Electrical Engineering from National Taiwan University of Science and Technology in 2019. Subsequently, he earned his M.S. degree from the Graduate Institute of Communication Engineering (GICE) at National Taiwan University in 2021. From 2021 to 2023, he was a system design engineer for WiFi chips at Realtek Semiconductor Corporation. From 2023 to 2025, he worked as a research assistant, first at the GICE, National Taiwan University, and later at the Department of Computer Science, National Yang Ming Chiao Tung University. He is currently a Ph.D. student in Electrical and Computer Engineering at the University of Michigan, Ann Arbor. His research interests include hypothesis testing, sequential inference, and statistical learning and control.
\end{IEEEbiographynophoto}

\begin{IEEEbiographynophoto}{I-Hsiang Wang}(Member, IEEE) received the B.Sc. degree in Electrical Engineering from National Taiwan University, Taiwan, in 2006. He received a Ph.D. degree in Electrical Engineering and Computer Sciences from the University of California at Berkeley, USA, in 2011. From 2011 to 2013, he was a postdoctoral researcher at \`{E}cole Polytechnique F\`{e}d\`{e}rale de Lausanne, Switzerland. Since 2013, he has been at the Department of Electrical Engineering in National Taiwan University, where he is now a professor. His research interests include network information theory, networked data analysis, and statistical learning. He was a finalist of the Best Student Paper Award of IEEE International Symposium on Information Theory, 2011. He received the 2017 IEEE Information Theory Society Taipei Chapter and IEEE Communications Society Taipei/Tainan Chapters Best Paper Award for Young Scholars.
\end{IEEEbiographynophoto}

\end{document}